\documentclass[preprint,10pt,review]{elsarticle}

\usepackage{lineno}
\usepackage{subcaption}
\usepackage{multirow}
\usepackage[table]{xcolor}

\usepackage[margin=2.5cm]{geometry}
\usepackage{hyperref}
\usepackage{amsmath,amsfonts,amssymb}
\usepackage{siunitx}
\usepackage{bm}
\usepackage{booktabs}
\usepackage{rotating} 

\DeclareMathOperator{\E}{\mathbb{E}}
\usepackage{mathtools}

\usepackage{relsize} 

\usepackage{algorithm}
\usepackage{algpseudocode}
\usepackage{wasysym}
\usepackage{paralist}

\newcommand{\Mod}[1]{\ (\mathrm{mod}\ #1)}
\usepackage{textcomp}

\biboptions{sort&compress} 

\begin{document}
\begin{frontmatter}

\title{An AI-based Domain-Decomposition Non-Intrusive Reduced-Order Model for Extended Domains applied to Multiphase Flow in Pipes}

\author[address_amcg]{Claire E. Heaney}
\author[address_amcg]{Zef Wolffs}
\author[address_amcg]{J{\'o}n Atli T{\'o}masson}
\author[address_chemeng]{Lyes Kahouadji}
\author[address_amcg]{Pablo Salinas}
\author[address_bp]{Andr{\'e} Nicolle}
\author[address_chemeng]{Omar K. Matar} 
\author[address_fsu]{Ionel M. Navon}
\author[address_ncl]{Narakorn Srinil}
\author[address_amcg]{Christopher C. Pain}

\address[address_amcg]{Applied Modelling and Computation Group, Department of Earth Science and Engineering, Imperial College London, SW7 2AZ}
\address[address_chemeng]{Department of Chemical Engineering, Imperial College London, Prince Consort Road, SW7 2AZ}
\address[address_bp]{AMT Group, BP Plc, Sunbury-upon-Thames} 
\address[address_fsu]{Department of Scientific Computing, Florida State University, Tallahassee, Florida 32306-4120, USA}
\address[address_ncl]{School of Engineering, Newcastle University, Newcastle upon Tyne, NE1 7RU}

\begin{abstract}
The modelling of multiphase flow in a pipe presents a significant challenge for high-resolution computational fluid dynamics (CFD) models due to the high aspect ratio (length over diameter) of the domain. In subsea applications, the pipe length can be several hundreds of kilometres versus a pipe diameter of just a few inches. Approximating CFD models in a low-dimensional space, reduced-order models have been shown to produce accurate results with a speed-up of orders of magnitude. In this paper, we present a new AI-based non-intrusive reduced-order model within a domain decomposition framework (AI-DDNIROM) which is capable of making predictions for domains significantly larger than the domain used in training. This is achieved by \begin{inparaenum}[(i)] \item using a domain decomposition approach; \item using dimensionality reduction to obtain a low-dimensional space in which to approximate the CFD model; \item training a neural network to make predictions for a single subdomain; and \item using an iteration-by-subdomain technique to converge the solution over the whole domain. \end{inparaenum} To find the low-dimensional space, we explore several types of autoencoder networks, known for their ability to compress information accurately and compactly. The performance of the autoencoders is assessed on two advection-dominated problems: flow past a cylinder and slug flow in a pipe. To make predictions in time, we exploit an adversarial network which aims to learn the distribution of the training data, in addition to learning the mapping between particular inputs and outputs. This type of network has shown the potential to produce realistic outputs. The whole framework is applied to multiphase slug flow in a horizontal pipe for which an AI-DDNIROM is trained on high-fidelity CFD simulations of a pipe of length \SI{10}{m} with an aspect ratio of 13:1, and tested by simulating the flow for a pipe of length \SI{98}{m} with an aspect ratio of almost 130:1. Statistics of the flows obtained from the CFD simulations are compared to those of the AI-DDNIROM  predictions to demonstrate the success of our approach. 
\end{abstract}


\end{frontmatter}

\section{Introduction}\label{sec:introduction}

Non-intrusive reduced-order modelling (NIROM) has been the subject of intense research activity over the last five years, largely due to the advances made in machine learning and the re-application of these techniques to reduced-order modelling. This paper takes a step towards demonstrating how non-intrusive reduced-order models can generalise, by training the model on one domain and deploying it on a much larger domain. The method outlined here could be extremely useful for the energy industry, for example, in which pipelines of the order of kilometres in lengths and inches in diameter are used for the subsea transportation of fluids. With such high aspect ratios, these pipes are too long to be modelled by high-resolution computational fluid dynamics (CFD) models alone. In this paper, we propose a non-intrusive reduced-order model based on autoencoders (for dimensionality reduction), an adversarial network (for prediction) and a domain decomposition approach. We investigate the performance of several autoencoders for dimensionality reduction using two test cases (flow past a cylinder and multiphase slug flow in a horizontal pipe). With the method that performs best, we go on to demonstrate the NIROM approach on multiphase slug flow in a horizontal pipe, training the networks on CFD data from a \SI{10}{m} pipe with an aspect ratio of 13:1 and making predictions for the flow within a \SI{98}{m} pipe with an aspect ratio of almost 130:1. In the following paragraphs we give some background on reduced-order modelling (ROM) and NIROM; on dimensionality reduction methods and the use of autoencoders; on prediction; domain decomposition methods and ROM; and lastly on multiphase flow. The final two paragraphs summarise the main contributions of the paper and describe the layout of the rest of the paper.

The aim of reduced-order modelling~\cite{Schilders2008} (ROM) is to obtain a low-dimensional approximation of a computationally expensive high-dimensional system of discretised equations, henceforth referred to as the high-fidelity model (HFM). To be of benefit, the low-dimensional model should be accurate enough for its intended purpose and orders of magnitude faster to solve than the HFM. Known as projection-based ROM~\cite{Benner2013}, one common strategy for constructing reduced-order models is to use a Galerkin projection of the HFM onto a low-dimensional subspace. However, in this article we focus on an alternative method, NIROM, which, unlike projection-based ROM, does not require access to or modification of the source code of the HFM. It requires only the results of the HFM with which it constructs a low-dimensional approximation to the HFM in two stages: the offline stage and the online stage. During the offline stage, solutions from the HFM are generated (known as snapshots); a set of basis functions that span the low-dimensional or reduced space are obtained by a dimensionality reduction method; and finally, the evolution of the HFM in the reduced space is approximated in some manner. This latter step can be done in several ways, but here, as we focus on AI-based NIROM, we use a neural network. 
A profusion of terms exist for this type of non-intrusive modelling, including POD with interpolation~\cite{Bui-Thanh2003}; NIROM~\cite{Audouze2013,Guenot2013}; POD surrogate modelling~\cite{Guenot2013,Hamdaoui2014}; system or model identification \cite{Polifke2014,Wang2017}; Galerkin-free~\cite{Shinde2016}; data-driven reduced-order modelling~\cite{Kaiser2014,Guo2019,Swischuk2019}; Deep Learning ROM~\cite{Fresca2020};  and digital twins~\cite{Rasheed2020,Kapteyn2020,AIAA2020,Niederer2021}. In addition to making predictions, digital twins assimilate data from observations to improve the  accuracy of the prediction. 

To find the low-dimensional subspace in which to approximate the HFM, many of these non-intrusive approaches rely on Proper Orthogonal Decomposition (POD)~\cite{Holmes2012} which is based on Singular Value Decomposition. Also known as Principal Component Analysis, POD finds the optimal linear subspace (with a given dimension) that can represent the space spanned by the snapshots and prioritises the modes according to those that exhibit the most variance. 
Whilst POD works well in many situations, for advection-dominated flows with their slow decay of singular values or large Kolmogorov $N$--width~\cite{Greif2019}, approximations based on POD can be poor~\cite{iollo2014advection,Ahmed2020b,Lu2020} and researchers are turning increasingly to autoencoders~\cite{Brunton2020}. Although adding to the offline cost, these networks seek a low-dimensional nonlinear subspace, which can be more accurate and efficient than a linear subspace for approximating the HFM. 

Convolutional networks are particularly good at analysing and classifying images (on structured grids)~\cite{Krizhevsky2012,He2015} with the ability to pick out features and patterns wherever their location (translational invariance), and these methods are applicable directly to the dimensionality reduction of CFD solutions on structured grids through the use of convolutional autoencoders (CAEs). Methods that apply convolutional networks to data on unstructured meshes do exist (based on space-filling curves~\cite{heaney2020applying}; graph convolutional networks~\cite{Hanocka2019,Tencer2021} and a method that introduces spatially varying kernels~\cite{Zhou2020NeurIPS}), but are in their infancy, so most researchers either solve the high-resolution problem on structured grids directly, or  interpolate from the high-fidelity model snapshots to a structured grid before applying the convolutional layers. The latter approach is adopted here. 

Perhaps the first use of an autoencoder for dimensionality reduction within a ROM framework was applied to reconstruct flow fields in the near-wall region of channel flow based on information at the wall~\cite{Milano2002}, whilst the first use of a convolutional autoencoder came 16 years later and was applied to Burgers Equation, advecting vortices and lid-driven cavity flow~\cite{gonzalez2018deep}. In the few years since 2018, many papers have appeared, in which convolutional autoencoders have been applied to sloshing waves, colliding bodies of fluid and smoke convection~\cite{Wiewel2019}; flow past a cylinder~\cite{fukami2020convolutional,Eivazi2020,Wu2021}; the Sod shock test and transient wake of a ship~\cite{Xu2020}; air pollution in an urban environment~\cite{Mack2020,Quilodran2020urban,Quilodran2021}; parametrised time-dependent problems~\cite{Nikolopoulos2021}; natural convection problems in porous media~\cite{Kadeethum2021porous}; the inviscid shallow water equations~\cite{Maulik2021rnncae}; supercritical flow around an airfoil~\cite{Wang2021}; cardiac electrophysiology~\cite{Fresca2021}; multiphase flow examples~\cite{Botsas2021}; the Kuramoto-Sivashinsky equation~\cite{Gin2021}; the parametrised 2D heat equation~\cite{Gruber2021}; and a collapsing water column~\cite{FuXiao2021}.  
Of these papers, those which compare autoencoder networks with POD generally conclude that autoencoders can outperform POD~\cite{gonzalez2018deep,fukami2020convolutional}, especially when small numbers of reduced variables are used~\cite{Kadeethum2021porous,Maulik2021rnncae,Wang2021,Fresca2021}. However, when large enough numbers of POD basis functions are retained, POD can yield good results, sometimes outperforming the autoencoders. 

A recent dimensionality reduction method that combines POD/SVD and an autoencoder (SVD-AE), has been introduced independently by a number of researchers and demonstrated on: vortex-induced vibrations of a flexible offshore riser at high Reynolds number~\cite{Reddy2019} (described as hybrid ROM); the generalised eigenvalue problems associated with neutron diffusion~\cite{phillips2020autoencoder} (described as an SVD autoencoder); Marsigli flow~\cite{Ahmed2020a} (described as nonlinear POD); and cardiac electrophysiology~\cite{Fresca2022} (described as POD-enhanced deep learning ROM). This method has at least three advantages: \begin{inparaenum}[(i)]\item by training the autoencoder with POD coefficients, it is of no consequence whether the snapshots are associated with a structured or unstructured mesh; \item an initial reduction of the number of variables by applying POD means that the autoencoder will have fewer trainable parameters and therefore be easier to train; and \item autoencoders in general can find the minimum number of latent variables needed in the reduced representation. For example, the solution of flow past a cylinder evolves on a one-dimensional manifold parametrised by time, therefore only one latent variable is needed to capture the physics of this solution~\cite{heaney2020applying,Maulik2021rnncae,Fresca2021}. \end{inparaenum} 

The Adversarial Autoencoder~\cite{makhzani2015adversarial} (AAE) is a generative autoencoder sharing similarities with the variational autoencoder (VAE) and the generative adversarial network (GAN). In addition to an encoder and decoder, the AAE has a discriminator network linked to its bottleneck layer. The purpose of the discriminator and associated adversarial training is to make the posterior distribution of the latent representation close to an arbitrary prior distribution thereby reducing the likelihood that the latent space will have `gaps'. Therefore, any set of latent variables should be associated, through the decoder, with a realistic output. Not many examples exist of using an AAE for dimensionality reduction in fluid dynamics problems, however, it has been applied to model air pollution in an urban environment~\cite{Quilodran2020urban,Quilodran2021}. In this work we compare POD, CAE, AAE and the SVD-AE on flow past a cylinder and multiphase flow in a pipe, to assess their suitability as dimension reduction methods.

Once the low-dimensional space has been found, the snapshots are projected onto this space, and the resulting reduced variables (either POD coefficients or latent variables of an autoencoder) can be used to train a neural network, which attempts to learn the evolution of the reduced variables in time (and/or their dependence on a set of parameters). From the references in this paper alone, many examples exist of feed-forward and recurrent neural networks having been used for the purpose of learning the evolution of time series data, for example, by Multi-layer perceptrons~\cite{Hesthaven2018,Raissi2019,Regazzoni2019,Swischuk2019,Pawar2019,Ahmed2020c,Fresca2020,Chen2021,Kadeethum2021porous,Nikolopoulos2021,Arthurs2021,Wang2021}, Gaussian Process Regression~\cite{Guo2019,Xiao2019CMAME,Xiao2019CAF,Maulik2021GPR,Botsas2021} and Long-Short Term Memory networks~\cite{gonzalez2018deep,Wiewel2019,Quilodran2020urban,Ahmed2020a,Eivazi2020,Maulik2021stable,Wu2021}. When using these types of neural network to predict in time, if the reduced variables stray outside of the range of values encountered during training, the neural network can produce unphysical, divergent results~\cite{Ahmed2020a,Quilodran2021ALSTM,Quilodran2021,Maulik2021stable,Fresca2022}. To combat this, a number of methods have been proposed. Physics-informed neural networks~\cite{Raissi2019} aim to constrain the predictions of the neural network to satisfy physical laws, such as conservation of mass or momentum~\cite{Arthurs2021,Chen2021}. A method introduced by References~\cite{Regazzoni2019,Pawar2019} aims to learn the mapping from the reduced variables at a particular time level to their time derivative, rather than the reduced values themselves at a future time level. This enables the use of variable time steps when needed, to control the accuracy of the solution in time. A third way of tackling this issue, which is explored in this paper, is to use adversarial networks, renowned for their ability to give realistic predictions. 

Adversarial networks, such as the GAN and the AAE, aim to learn a distribution to which the training data could belong, in addition to a mapping between solutions at successive time levels. GANs and AAEs are similar in that they both use a discriminator network and deploy adversarial training, and both require some modification so that they can make predictions in time. The aim of these networks is to generate images (or in this case, reduced variables associated with fluid flows) that are as realistic as possible. To date there are not many examples of the use of GANs or AAEs for prediction in CFD modelling. Two exceptions are Reference~\cite{Cheng2020AAE} which combines a VAE and GAN to model flow past a cylinder and the collapse of a water dam; and Reference~\cite{Silva2021DA} which uses a GAN to predict the reduced variables of an epidemiological model which modelled the spread of a virus through a small, idealised town. This particular model performed well when compared with an LSTM~\cite{quilodrancasas2021digital}. Conditional GANs (CGAN) have similar properties to the GAN and AAE, and they have been used successfully to model forward and inverse problems for coupled hydro-mechanical processes in heterogeneous porous media~\cite{Kadeethum2021CGAN}; a flooding event in Hokkaido, Japan, after the 1993 earthquake~\cite{Cheng2020DCGAN}; and a flooding event in Denmark~\cite{Cheng2021}.  However, the closeness of the CGAN's distribution to that of the training data is compromised by the `condition' or constraint.  GANs are known to be difficult to train, so, in this paper, we use an Adversarial Autoencoder, albeit modified, so that it can predict the evolution of the reduced variables in time.   

Combining domain decomposition techniques with ROM has been done by a number of researchers. An early example~\cite{Baiges2013} presents a method for projection-based ROMs in which the POD basis functions are restricted to the nodes of each subdomain of the partitioned domain. A similar approach has also been developed for non-intrusive ROMs~\cite{Xiao2019CMAME}, which was later extended to partition the domain by minimising communication between subdomains~\cite{Xiao2019CAF}, effectively isolating as much as possible, the physical complexities between subdomains. As the domain of our main test case (multiphase flow in a pipe) is long and thin with a similar amount of resolution and complexity of behaviour occurring in partitions of equal length in the axial direction, here, we simply split the domain into subdomains of equal length in the axial direction (see Figure~\ref{fig:schematic_dom_decomp}). The neural network learns how to predict the solution for a given subdomain, and the solution throughout the entire pipe is built up by using the iteration-by-subdomain approach~\cite{Gastaldi1992}. The domain decomposition approach we use has some similarities to the method employed in Reference~\cite{Yang2021}, which decomposes a domain into patches to make training a neural network more tractable. However, our motivation for using domain decomposition is to make predictions for domains that are significantly larger than those used in the training process. When modelling a pipe that is longer than the pipe used to generate the training data, it is likely that the simulation will need to be run for longer than the original model as the fluid will take longer to reach the end of the pipe. This means that boundary conditions for the longer pipe must be generated somehow, rather than relying on using boundary conditions from the original model. Generating suitable boundary conditions for turbulent CFD problems is, in general, an open area of research. Often used are incoming synthetic-eddy methods~\cite{Skillen2016}, which attempt to match specified mean flows and Reynolds stresses at the inlet. Recently, researchers have explored using GANs to generate boundary conditions with success~\cite{Fukami2019,Kim2020inlet}. We present three methods of generating boundary conditions for our particular application and also discuss alternative methods in Conclusions and Further Work.  


The test case of multiphase flow in a pipe is particularly challenging due to the difficulties such as the space-time evolution of multiphase flow patterns (stratified, bubbly, slug, annular), the turbulent phase-to-phase interactions, the drag, inertia and wake effects that arise for the HFM from the high aspect ratio (length to diameter) of the domain of a typical pipe. Many address this by developing one dimensional (flow regime-dependent or -independent) models for long pipes~\cite{kjolaas2013simulation,bonzanini2017simplified,KRASNOPOLSKY2018}. Nevertheless, such models contain some uncertainties as they rely on several closure or empirical expressions~\cite{Ma2020} under the limited experimental data~\cite{KIM2020} in describing, for example, the 3D space-time variations of interfacial frictional forces with phase distributions (the bubble/drop entrainment, the bubble-induced turbulence, the phase interfacial interactions), depending on the flow pattern, flow direction and pipe physical properties (inclination, diameter, length). Significant progress has been made in 3D modelling~\cite{Tryggvason2020conf} by using direct numerical simulations~\cite{Xie2020} (DNS) and front-tracking methods~\cite{Tryggvason2011book}. To generate the solutions of the HFM, we employ a method based on Large Eddy Simulation, which advects a volume fraction field~\cite{Obeysekara2021} and uses mesh adaptivity to have high resolution where most needed. Although compromising on resolving features on the smaller temporal and spatial scales, this approach is computationally more feasible than DNS and has the advantage of being conservative, unlike front-tracking methods. 

In this paper, we propose a non-intrusive reduced-order model (AI-DDNIROM) capable of making predictions for a domain to which it has not been exposed during training. Several autoencoders are explored for the dimensionality reduction stage, as there is evidence that they are more efficient than POD for advection-dominated problems such as those tackled here. 
The dimensionality reduction methods are applied to 2D flow past a cylinder and 3D multiphase slug flow in a horizontal pipe. For the prediction stage, an adversarial network is chosen (based on a modified adversarial autoencoder) as these types of networks are believed to generate latent spaces with no gaps~\cite{makhzani2015adversarial} and thus are likely to produce more realistic results than feed-forward or recurrent neural networks without adversarial layers. A domain decomposition approach is applied, which, with an iteration-by-subdomain technique, enables predictions to be made for multiphase slug flow with a significantly longer pipe than was used when training the networks. Statistics from the HFM solutions, and predictions of the non-intrusive reduced-order models for the original length pipe and the longer pipe are compared. The contributions of this work are: \begin{inparaenum}[(i)] 
    \item a method which can make predictions for a domain significantly larger than that used to train the reduced-order models; 
    \item the exploitation of an adversarial network to make realistic predictions, and comparing statistics of the reduced-order models with the original CFD model; 
    \item the an investigation of a number of methods to generate boundary conditions for the larger domain. 
\end{inparaenum}

The outline of the remainder of the paper is as follows. Section~\ref{sec:methodology} describes the methods used in constructing the reduced-order models and the domain decomposition approach which is exploited in order to be able to make predictions for a longer domain than that used in training. Section~\ref{sec:results} presents the results for the dimensionality reduction methods applied to flow past a cylinder and multiphase flow in a pipe, and then shows the predictions of the reduced-order model of multiphase flow in a pipe, for both the original domain and the extended domain. Conclusions are drawn and future work described in the final section. Details of the hyperparameter optimisation process and the network architectures are given in the appendix.

\section{Methodology}\label{sec:methodology}
\subsection{Non-intrusive reduced-order models}
The offline stage of a non-intrusive reduced-order model can be split into three steps: \begin{inparaenum}[(i)]\item generating the snapshots by solving a set of discretised governing equations (the high-resolution or high-fidelity model); \item reducing the dimensionality of the discretised system; and \item teaching a neural network to predict the evolution of the snapshots in reduced space. \end{inparaenum} The online stage consists of two steps: \begin{inparaenum}[(i)] \item predicting values of the reduced variables with the neural network for an unseen state; and \item mapping back to the physical space of the high-resolution model. \end{inparaenum} %
In this section, the methods used in this investigation for dimensionality reduction (Section~\ref{sec:compression}) and prediction (Section~\ref{sec:prediction}) are described. The final section (Section~\ref{sec:extension}) outlines an approach for making predictions for a larger domain having used a smaller domain to generate the training data.  

\subsection{Dimensionality reduction methods}\label{sec:compression}
Described here are four techniques for dimensionality reduction which are used in this investigation, namely Proper Orthogonal Decomposition, a convolutional autoencoder, an adversarial autoencoder and a hybrid SVD autoencoder. 

\subsubsection{Proper Orthogonal Decomposition}
Proper Orthogonal Decomposition is a commonly used technique for dimensionality reduction when constructing reduced-order models. POD requires the minimisation of the reconstruction error of the projection of a set of solutions (snapshots) onto a number of basis functions which define a low-dimensional space. In order to minimise the reconstruction error, the basis functions must be chosen as the left singular vectors of the singular value decomposition (SVD) of the matrix of snapshots. Suppose the snapshots matrix is represented by $\bm{S}$, whose columns are solutions at different instances in time (i.e.~the snapshots) and whose rows correspond to nodal values of solution variables, then $\bm{S}$ can be decomposed as 
\begin{equation}
    \bm{S} = \bm{U} \Sigma \bm{V}^T \,,
\end{equation}
where the matrix $\bm{U}$ contains the left singular vectors, $\bm{V}$ the right singular vectors and $\Sigma$ contains the singular values on its diagonal, zeros elsewhere. If POD is well suited to the problem, many of the singular values will be close to zero and the corresponding columns of $\bm{U}$ can be discarded. The POD basis functions to be retained are stored in a matrix denoted by $\bm{R}$. The POD coefficients of a snapshot can be found by pre-multiplying the snapshot by $\bm{R}^T$, and the reconstruction of a snapshot can be found by pre-multiplying the POD coefficients of the snapshot by $\bm{R}$:
\begin{equation}\label{eq:recon_pod}
    (\bm{u}^{\text{recon}})^k = \bm{R}\bm{R}^T\bm{u}^k\,,
\end{equation}
where $\bm{u}^k$ is the $k$th snapshot and  $(\bm{u}^{\text{recon}})^k$ is its reconstruction. Hence the reconstruction error over a set of $N$~snapshots $\{\bm{u}^1, \bm{u}^2, \ldots, \bm{u}^N\}$ can be written as
\begin{equation}\label{eq:recon_error_mse}
   \frac{1}{N}\sum_{k=1}^{N}\left(\bm{u}^k - (\bm{u}^{\text{recon}})^k\right) \bm{\cdot} \left(\bm{u}^k - (\bm{u}^{\text{recon}})^k\right)\,.
\end{equation}
Often the mean is subtracted from the snapshots before applying singular value decomposition, however, in this study, doing so was found to have little effect. In the first test case, 2D flow past a cylinder, two velocity components are included in the snapshots matrix:
\begin{equation}
    \bm{u}^k = (u_1^k, u_2^k, \ldots, u_M^k, v_1^k, v_2^k, \ldots, v_M^k)^T
\end{equation}
where $u_i$ and $v_i$ represent the $x$ and $y$ components of velocity, respectively, at the $i$th node; $k$ denotes a particular snapshot; and $M$ is the number of nodes. For the 3D multiphase flow test case the snapshots comprise velocities and volume fractions, so a single snapshot has the form
\begin{equation}
    \bm{u}^k = (u_1^k, u_2^k, \ldots, u_M^k, v_1^k, v_2^k, \ldots, v_M^k, w_1^k, w_2^k, \ldots, w_M^k, \alpha_1^k, \alpha_2^k, \ldots, \alpha_M^k)^T
\end{equation}
where $w_i^k$ and  $\alpha_i^k$ represent the $z$ component of velocity and the volume fraction, respectively, at the $i$th node of the $k$th snapshot. In this case, the velocity components are scaled to be in the range $[-1,1]$ so that their magnitudes are similar to those of the volume fractions.

\subsubsection{Convolutional Autoencoder}
An autoencoder is a particular type of feed-forward network that attempts to learn the identity map~\cite{Baldi1989}. When used for compression, these networks have a central or bottleneck layer that has fewer neurons than the input and output layers, thereby forcing the autoencoder to learn a compressed representation of the training data. An autoencoder consists of an encoder which compresses the data to the latent variables of the bottleneck layer, and a decoder which decompresses or reconstructs the latent variables to an output layer of the same dimension as the input layer. The latent variables span what is referred to as the latent space.  
The convolutional autoencoder typically uses a series of two types of layers to compress the input data in the encoder: convolutional layers and pooling layers. These layers both apply operations to an input grid resulting in an output grid (or feature map) of reduced size. The inverse operations are then used in succession in a decoder, resulting in a reconstructed grid of the same shape as the input. The encoder-decoder pair can be trained as any other neural network:  by passing training data through the network and updating the weights associated with the layers according to a loss function such as the mean square error. If $\bm{u}^k$ represents the $k$th sample in the dataset of $N$ samples and $(\bm{u}^{\text{recon}})^k$ represents the corresponding output of the autoencoder, which can be written as
\begin{equation}\label{eq:recon_ae}
(\bm{u}^{\text{recon}})^k = f^{\text{ae}}(\bm{u}^k)\,,    
\end{equation} 
then the mean square error can be expressed as in Equation~\eqref{eq:recon_error_mse}.

\subsubsection{Adversarial Autoencoder}
\label{section:aae}
The adversarial autoencoder \cite{makhzani2015adversarial} is a recently developed neural network that uses an adversarial strategy to force the latent space to follow a (given) prior distribution ($P_{\text{prior}}$). Its encoder-decoder network is the same as that of a standard autoencoder, however, in addition, the adversarial autoencoder includes a discriminator network, which is trained to distinguish between true samples (from the prior) and fake samples (from the latent space). %
There are therefore three separate training steps per mini-batch. In the first step  the reconstruction error of the inputs is minimised (as is done in a standard autoencoder). In the second and third steps, the adversarial training takes place.  In the second step, the discriminator network is trained on latent variables sampled from the prior distribution with label~1 and latent variables generated by the encoder with label~0. In the third step, the encoder is trained to fool the discriminator, that is, it tries to make the discriminator produce an output of~1 from its generated latent vectors. Note that this is the role of the generator in a GAN and, as such, the encoder ($G$) and discriminator ($D$) play the minimax game described by Equation~\eqref{eq:minimax}. This equation is the implicit loss function for the adversarial training:
\begin{equation}
\label{eq:minimax}
    \min_{G} \max_{D} V (D, G) = \E_{\bm{z} \sim P_{\text{prior}}}[\log D(\bm{z})]  + \E_{\bm{u} \sim P_{\text{data}}}[\log(1- D(G(\bm{u})))]\ ,
\end{equation}
where $V$ is the value function that $G$ and $D$ play the minimax game over, $\bm{z} \sim P_\text{prior}$ is a sample from the desired distribution and $\bm{u} \sim P_{\text{data}}$ is a sample input grid. There are strong similarities between the adversarial autoencoder, GANs and Variational Autoencoders (VAEs). All three types of network set out to obtain better generalisation than non-adversarial networks by attempting to obtain a smooth latent space with no gaps. 
Results in Reference~\cite{makhzani2015adversarial} show that the AAE performs better at this task than the VAE on the MNIST digits. 
Imposing a prior distribution upon the variables of the latent space ensures that any set of latent variables, when passed through the decoder, should have a realistic  output~\cite{makhzani2015adversarial}.

\subsubsection{SVD Autoencoder}\label{app:svd-ae}
As the name suggests, the SVD autoencoder makes use of two strategies. Initially an SVD is applied to the data, resulting in POD coefficients that are subsequently used to train an autoencoder, which applies a second level of compression. Once trained, the latent variables of the SVD autoencoder can be written as 
\begin{equation}
    \bm{z}^k = f^{\text{enc}}\left( \bm{R}\bm{u}^k\right)
\end{equation}
where $f^{\text{enc}}$ is the encoder, $\bm{R}$ represents the POD basis functions, $\bm{u}^k$ is the $k$th snapshot and $\bm{z}^k$ are the latent variables. For reconstruction,  the inverse of this process is then employed, whereby a trained decoder first decompresses the latent space variables to POD coefficients, after which these POD coefficients are reconstructed to the original space of the high-fidelity model.  The reconstruction can be written as 
\begin{eqnarray}\label{eq:recon_svdae}
    (\bm{u}^{\text{recon}})^k &=& \bm{R}^T f^{\text{dec}}\left(f^{\text{enc}}\left( \bm{R}\bm{u}^k\right)\right) \equiv \bm{R}^T f^{\text{ae}}\left( \bm{R}\bm{u}^k\right)\,,
\end{eqnarray}
where $f^{\text{dec}}$ is the decoder, $f^{\text{ae}}$ is the autoencoder and $(\bm{u}^{\text{recon}})^k$ is the reconstruction of the $k$th snapshot. This network could be approximated by adding to the autoencoder a linear layer after the input and before the output, and dispensing with the SVD, however, it has been found that this network is harder to train. Here, we take advantage of the efficiency of the SVD and use this in conjunction with an autoencoder.

\subsection{Prediction methods}\label{sec:prediction}

In this study, when predicting, we wish to approximate a set of reduced variables (either POD coefficients or latent variables of an autoencoder) at a future timestep. The adversarial autoencoder is re-purposed for this task in an attempt to capitalise on the fact that this network should produce realistic results (providing that the training data is representative of the behaviour that will be modelled). So that it can predict time series data, three modifications are made to the original adversarial autoencoder network~\cite{makhzani2015adversarial}: namely that \begin{inparaenum}[(i)] \item the bottleneck layer no longer has fewer variables than the input (to prevent further compression); \item the output is the network's approximation of the reduced variables at a future time level; and \item the input is the reduced variables at the preceding time level as well as the reduced variables of the neighbouring subdomains at the future time (as we adopt a domain decomposition approach which is described in the next paragraph). \end{inparaenum} The modified adversarial autoencoder is trained by minimising the error between its output and the predicted variables at the future time level, as well as incorporating the adversarial training strategy described in Section~\ref{section:aae}. To avoid confusion, we refer to this network as a predictive adversarial network, because, with different inputs and outputs, it is no longer an autoencoder. 

\begin{figure}[htbp]
\centering
\includegraphics[width=8.5cm,trim=0mm 0mm 0mm 28mm, clip]{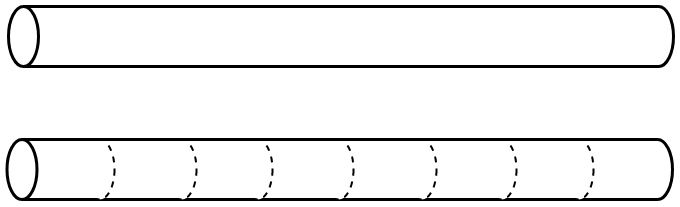}
\caption{A schematic diagram of a pipe split into 8 subdomains of equal length in the axial direction}
\label{fig:schematic_dom_decomp}
\end{figure}

In this study we adopt a domain decomposition approach to facilitate predicting the solution for larger domains than that used in training (see next section). Given the aspect ratio of the pipe, we split the domain into subdomains of equal length in the axial direction, see Figure~\ref{fig:schematic_dom_decomp}. To train the predictive adversarial network, reduced variables are obtained by interpolating the high-fidelity solutions or snapshots onto a structured grid in each subdomain in turn and compressing the interpolated snapshots from all the subdomains using POD or an autoencoder. The interpolation is linear and achieved by using the finite element basis functions. The predictive adversarial network is taught to predict the reduced variables in a particular subdomain at a future time level given the reduced variables in the neighbouring subdomains at the future time level and the reduced variables in the subdomain at the preceding time level. Using training data for all the subdomains and those time levels that are in the training dataset, the predictive adversarial network learns the mapping~$f$, written as
\begin{equation}\label{eq:predictive_adversarial}
    \bm{z}_i^k = f(\bm{z}_{i-1}^k, \,\bm{z}_i^{k-1},\, \bm{z}_{i+1}^k)\,, \qquad \forall i
\end{equation}
where $\bm{z}_i^k$ represents the reduced variables in subdomain~$i$ at the future time level~$k$; $\bm{z}_i^{k-1}$ represents the same but at the preceding time level; and $\bm{z}_{i-1}^k$ and $\bm{z}_{i+1}^k$ denote the reduced variables at the future time level for the subdomains to the left and right of subdomain~$i$. When predicting for one time level, all subdomains are iterated over (the iteration-by-subdomain method) until convergence is reached over the whole domain. This is done by sweeping from left to right (increasing~$i$) and then sweeping from right to left (deceasing~$i$). During the iteration process, $\bm{z}^k_i$ of Equation~\eqref{eq:predictive_adversarial} is being continually updated. As we consider incompressible flows in this study, the solution method has to be implicit in order to allow information to travel throughout the domain within one time level. This sweeping from left to right and back again allows information to pass from the leftmost to the rightmost subdomains and vice versa.

For each new time level, an initial solution is required to start the iteration process (for $\bm{z}_{i-1}^{k}$ and $\bm{z}_{i+1}^{k}$ in Equation~\eqref{eq:predictive_adversarial}). The solution at the previous, converged time level could be used ($\bm{z}_{i\pm1}^{k}=\bm{z}_{i\pm1}^{k-1}$), however, using linear extrapolation based on two previous time levels showed better convergence:
\begin{equation}
    \bm{z}^{k}_i = \bm{z}^{k-1}_i + (\bm{z}^{k-1}_i - \bm{z}^{k-2}_i)\,. 
\end{equation}

The procedure for sweeping over the subdomains is given in Algorithm~\ref{alg:sampling}, in which $f$ represents the predictive adversarial network, $N^{\text{time}}$ is the number of time levels, $N^{\text{sweep}}$ is the number of sweeps carried out over the whole domain and $N^{\text{sub}}$ is the total number of subdomains. Two of these subdomains are treated as boundary conditions and are fully imposed throughout the duration of the prediction, so at line~\ref{alg:subdomain_loop} of Algorithm~\ref{alg:sampling}, only the subdomains where a solution is sought are iterated over. In this study, a fixed number of sweep iterations were used as this gave good results, however, a convergence criterion could be easily implemented if desired.

\begin{algorithm}[H]
\caption{\label{alg:sampling}An algorithm for finding the solution for the reduced variables in a subdomain and sweeping over all the subdomains to obtain a converged solution over the whole domain.}
\begin{algorithmic}[1]
\State \textit{!! set initial conditions for each subdomain~$i$}
\State $\bm{z}^0_i\  \forall i$
\For{time level $k=1,2,\ldots,N^{\text{time}}$} \label{alg:time_level_loop}
\State \textit{!! set boundary conditions} 
\State {$\bm{z}^k_1$, $\bm{z}^k_{N^{\text{sub}}}$}
\State \textit{!! estimate the solution at the future time level $k$ for all the subdomains}
\If{$k>1$}
\For{subdomain $i = 2,3,\ldots, N^{\text{sub}}-1$}
\State $\bm{z}^{k}_i = \bm{z}^{k-1}_i + (\bm{z}^{k-1}_i - \bm{z}^{k-2}_i)$
\EndFor
\Else
\For{subdomain $i = 2,3,\ldots, N^{\text{sub}}-1$}
\State $\bm{z}^{k}_i = \bm{z}^{k-1}_i$
\EndFor
\EndIf
\State \textit{!! sweep over subdomains}
\For{sweep iteration $j=1,2,\ldots, N^{\text{sweep}}$} 
\For{subdomain $i = 2,3,\ldots, N^{\text{sub}}-2,N^{\text{sub}}-1,N^{\text{sub}}-2,\ldots,4,3$} \label{alg:subdomain_loop}
\State \textit{!! calculate the latent variables of subdomain $i$ at time level $k$}
\State $\bm{z}^{k}_i = f(\bm{z}_{i-1}^{k},\, \bm{z}_i^{k-1}\!,\, \bm{z}_{i+1}^{k})$
\EndFor 
\EndFor 
\EndFor 
\end{algorithmic}
\end{algorithm}

\subsection{Extending the domain}\label{sec:extension}
In this study, we investigate the ability of a non-intrusive reduced-order model in combination with a domain decomposition approach to be able to make predictions for domains larger than that used in the training process. We test this approach on the dataset generated from multiphase flow in a pipe. With sufficient initial conditions and boundary conditions, exactly the same procedure can be used to make predictions for the extended domain as is used to make predictions for the domain used in training. That is, the solution is obtained for a single subdomain, whilst sweeping over all subdomains until convergence is reached (outlined in Section~\ref{sec:prediction}). 

As the length of the pipe of interest (`extended pipe') is longer than the pipe used in training, initial conditions must be generated throughout the extended pipe. The method used here is to specify initial conditions throughout the extended pipe by repeating initial conditions from the shorter pipe. An alternative would be to find the reduced variables for a steady state (for example, water in the bottom half of the pipe and air in the top half) and use these values in every subdomain in the extended pipe. We choose the former method to reduce the time taken for instabilities and slugs to develop. 

For the extended pipe, boundary conditions (effectively the reduced variables in an entire subdomain) can be imposed using the data already available from the HFM. However, as the length of the pipe is longer than the pipe used in training, we wish to make predictions over a longer period over which snapshots were collected from the HFM.  In order to obtain boundary conditions for the extended pipe, several methods are explored for the inlet or upstream boundary. Of those investigated, three methods performed better than the others, listed below.  
\begin{enumerate}[(i)]
    \item Cycling through slug formation: a slug is found in the shorter pipe, and the velocity and volume fraction fields associated with the advection of the slug through a subdomain are looped over in the upstream boundary subdomain. 
    \item Perturbed instability: the volume fraction field associated with an instability from the shorter pipe is perturbed with Gaussian noise. This is then imposed on the boundary subdomain. The associated velocity field is used unperturbed.
    \item Original boundaries repeated: solutions from the shorter pipe are cycled through in the boundary subdomain.
\end{enumerate}
At the downstream boundary, reduced variables corresponding to a steady state solution (water in the bottom half of the pipe and air in the top half) was imposed. Specific details of the boundary conditions are given in the results section. These three approaches are somewhat heuristic. As we are using information that the model will not have seen and that does not accurately satisfy the governing equations, we exploit the ability of the predictive adversarial network to produce realistic results, as it should have learnt appropriate spatial and temporal covariance information during training. An alternative method for generating boundary conditions is discussed in the section on conclusions and future work.

\section{Results}\label{sec:results}

\subsection{Test Cases}
Two test cases are used to demonstrate the dimensionality reduction methods proposed in this paper. The first is flow past a cylinder in 2D, the second is 3D multiphase flow in a pipe. The second test case is also used to demonstrate the prediction capabilities of the predictive adversarial network for both the domain that was used in training and a domain that is significantly longer that the one used in training. The test cases are now described.
\subsubsection{Flow past a cylinder}
The following partial differential equations describe the motion of an incompressible fluid: 
\begin{eqnarray}
 \bm{\nabla} \cdot \bm{u} & = & 0\,, \label{conmass} \\
\rho\left( 
\frac{\partial \bm{u}}{\partial t} + \bm{u}\cdot\bm{\nabla} \bm{u}
\right) 
-\bm{\nabla}\cdot\bm{\tau} & =& -\bm{\nabla}p\,,
\label{conmom}
\end{eqnarray}
where $\rho$ is the density (assumed constant), $\bm{u}$~is the velocity vector, $\bm{\tau}$~contains the viscous terms associated with an isotropic Newtonian fluid, $p$~represents the non-hydrostatic pressure, $t$~is time and the gradient operator $\bm{\nabla}$ is defined as 
\begin{equation}
\bm{\nabla} = \left(\frac{\partial}{\partial x}\,,\ \frac{\partial}{\partial y}\right)^T \,.
\end{equation}
When solving these equations, a linear triangular element is used with a discontinuous Galerkin discretisation for the velocities and a continuous Galerkin representation of the pressure (the P1DG-P1 element). To discretise in time, Crank-Nicolson is used. As the velocity field fully describes incompressible flow, only the velocity variables are required by the reduced-order models. For more details on how this system of equations is discretised and solved, the reader is referred to~\cite{Xie2016}. For the flow past a cylinder test case, the domain measures \SI{2.2}{m} (horizontal axis) by  \SI{0.41}{m} (vertical axis), and the centre of the cylinder is located at \SI{0.2}{m} from the leftmost boundary on the horizontal centreline of the domain. Free slip and no normal flow boundary conditions are applied on the upper and lower walls, no slip is applied on the surface of the cylinder. Zero shear and zero normal stress are applied at the outlet (the right-hand boundary of the domain). In the following results, speeds and velocities are given in metres per second and time is in seconds. A Reynolds number of 3900 was used:
\begin{equation}
    Re=\frac{\rho\, U L }{\mu}=3900\,,
\end{equation} 
where $U$ is the constant inlet velocity, $U=\SI{0.039}{\metre\per\second}$, the density has value $\rho=\SI{1000}{\kilogram\per\cubic\metre}$ and the diameter of the cylinder is $L=\SI{0.1}{m}$. Thus the dynamic viscosity is $\mu=\SI{e-3}{\kilo\gram\per\metre\per\second}$. Formed from solutions of this problem, the dataset consists of 2000 snapshots with a time interval of \SI{0.25}{s}. (An adaptive time-step was used to solve the equations, however the solutions were saved every \SI{0.25}{s} to generate the snapshots.) 


\subsubsection{Multiphase flow in a pipe}
Multiphase slug flow in a horizontal pipe is used as the second test case. We use an interface capturing method, in which we track the interface by solving an advection equation for the volume fraction of the liquid phase. Let $\alpha$ be the volume fraction of the liquid (water in this case), which means that the volume fraction of the gas (air) is $(1-\alpha)$. The conservation of mass for incompressible fluids can therefore be written as 
\begin{eqnarray}
\frac{\partial{}}{\partial{t}}({\alpha}) + \bm{\nabla}\cdot(\alpha\bm{u}) & = & 0 ,\\ 
\bm{\nabla}\cdot\bm{u} & = & 0 ,
\label{mass_conservation}
\end{eqnarray}
where $t$ represents time and $\bm{u}$ represents velocity. Assuming incompressible viscous fluids, conservation of momentum yields the following
\begin{equation}
\rho\left(\frac{\partial{\bm{u}}}{\partial{t}} + \bm{u}\cdot\bm{\nabla}\bm{u} \right) = -{\bm{\nabla}}p + \bm{\nabla} \cdot\left(\mu(\bm{\nabla}\bm{u}+\bm{\nabla}^T\bm{u})\right)+\rho\bm{g}+\bm{F}_\sigma
\,,
\label{incompress_NS}
\end{equation}
where $p$ represents pressure, $\bm{g}$ is the gravitational acceleration vector $(0,0,9.8)$ and $\bm{F}_\sigma$ is the force representing surface tension. The bulk density and bulk viscosity are defined as
\begin{eqnarray}
\rho& =& \alpha\rho_{\text{water}}+(1-\alpha)\rho_{\text{air}} , \\
\mu& =& \alpha\mu_{\text{water}}+(1-\alpha)\mu_{\text{air}} , 
\end{eqnarray}
respectively, where $\rho_{\text{water}}$ and $\rho_{\text{air}}$ are the densities of water and air respectively, and $\mu_{\text{water}}$ and $\mu_{\text{air}}$ are the dynamic viscosities of water and air respectively. 
For more details of how the governing equations are discretised and solved, see Reference~\cite{Obeysekara2021}, including information on the unstructured adaptive meshing process, the adaptive time stepping and compressive advection technique to keep the interface at the boundary of the fluids sharp. The densities of air and water are taken as \SI{1.125}{\kilo\gram\per\cubic\metre} and \SI{1000}{\kilo\gram\per\cubic\metre} respectively, and the viscosities are \SI{1.81e-05}{\kilo\gram\per\metre\per\second} and \SI{9.892e-04}{\kilo\gram\per\metre\per\second} respectively. The modelled pipe has dimensions of \SI{10}{m} in length and a radius of \SI{0.039}{m}. Boundary conditions of no normal flow and no slip were weakly enforced on the pipe wall, and any incoming momentum is given a value of zero. The outlet of the pipe has a non-hydrostatic pressure of zero, and again, any incoming velocities are set to zero and incoming volume fraction is taken to be water. Initially the pipe is filled entirely with water, which flows along the axial direction at a velocity of \SI{4.162}{\metre\per\second} in the top half of the pipe and \SI{2.082}{\metre\per\second} in the bottom half. After the first time-step, air starts flowing in through the inlet through the top half at a velocity of \SI{4.162}{\metre\per\second}, a scenario which can lead to the formation of slugs. These values of velocity correspond to superficial velocities of air and water of \SI{2.081}{\metre\per\second} and \SI{1.041}{\metre\per\second} respectively. 
The dataset used for training the reduced-order models consists of solutions at 800 time levels with a fixed time interval of \SI{0.01}{s}. The reduced-order models use the velocity fields in three directions and the volume fraction field.

\subsection{Dimensionality reduction}
\begin{sidewaysfigure*}[htbp]
\centering
\includegraphics[width=0.9\textwidth]{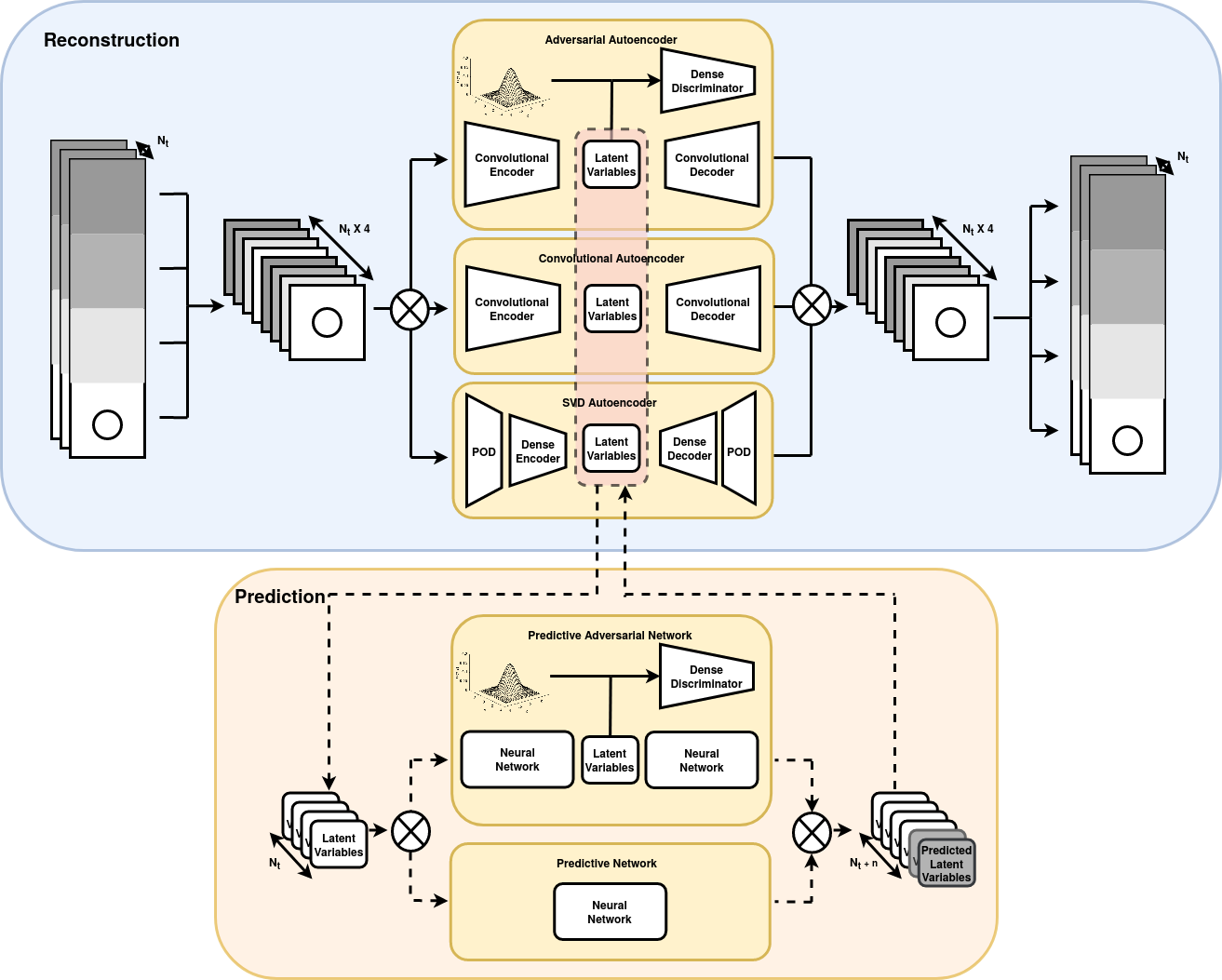}
\caption{Upper part (shaded blue): shows the training of the autoencoders for the dimensionality reduction of flow past a cylinder. Lower part (shaded orange): shows how the the predictive adversarial network was trained.}
\label{fig:schematic_workflow}
\end{sidewaysfigure*}

Four methods for dimensionality reduction (or compression) are compared, namely POD, CAE, AAE and SVD-AE.  An extensive hyperparameter optimsation was performed to find the optimal set of values for the hyperparameters of each autoencoder. Details of the hyperparameters that were varied, the ranges over which they were varied, and the optimal values and architectures that were obtained as a result can be found in Tables~\ref{tab:app_hpo}, \ref{tab:app_optimal_ae} and~\ref{tab:architecture_DR}. Ten POD basis functions were retained for the compression based on POD and ten latent variables were used in the bottleneck layers of the autoencoders. For the SVD-AE, one hundred POD coefficients were retained which were then compressed to ten latent variables by an autoencoder. The top part (shaded blue) of  Figure~\ref{fig:schematic_workflow} shows a schematic diagram of how the networks used for dimensionality reduction are trained for the flow past a cylinder test case. 


\subsubsection{Flow past a cylinder}
The CFD solutions were saved every \SI{0.25}{\second} for \SI{500}{\second} resulting in 2000 snapshots. The domain was split into four subdomains, each spanning the entire height of the domain and a quarter of its length. These were discretised with 20~by~20 structured grids. The velocity solutions from the unstructured mesh were linearly interpolated onto the four grids using the finite element basis functions, resulting in a dataset of 8000 samples. For POD, the columns of the snapshots matrix consisted of values of both velocity components, and for the autoencoders, the two velocity components were fed into two separate channels. The training data (which also includes the validation data) was formed by randomly selecting 7200 samples from the full dataset. The remaining 800 samples were used as the test dataset (i.e.~unseen data). 

\begin{figure*}[htbp]
    \centering
    \includegraphics[width=0.45\textwidth]{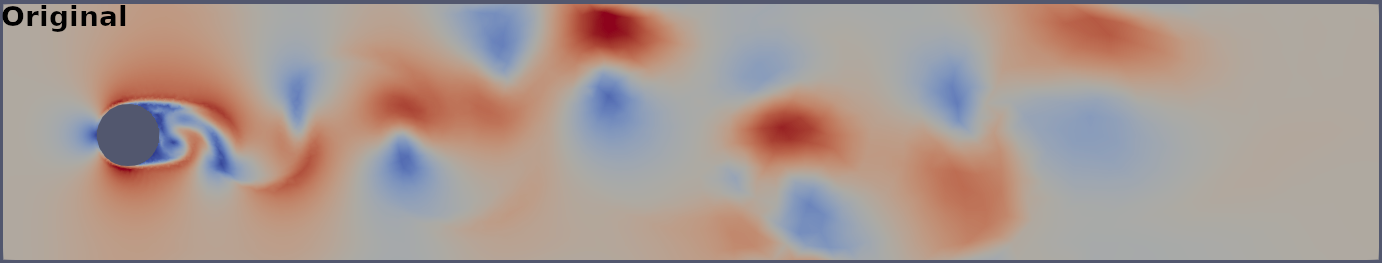}
    \includegraphics[width=0.45\textwidth]{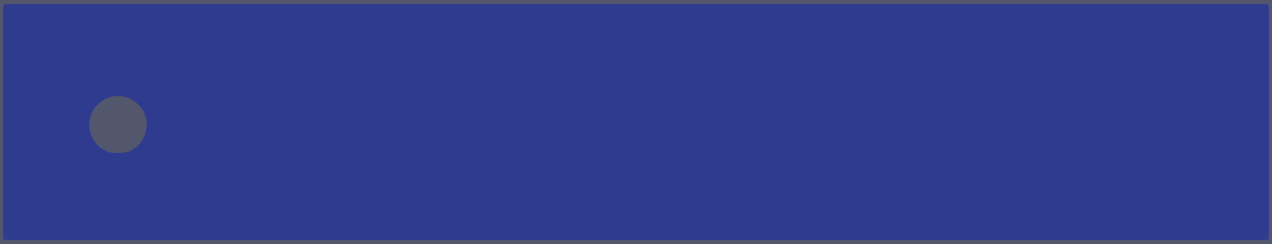}
    \includegraphics[width=0.45\textwidth]{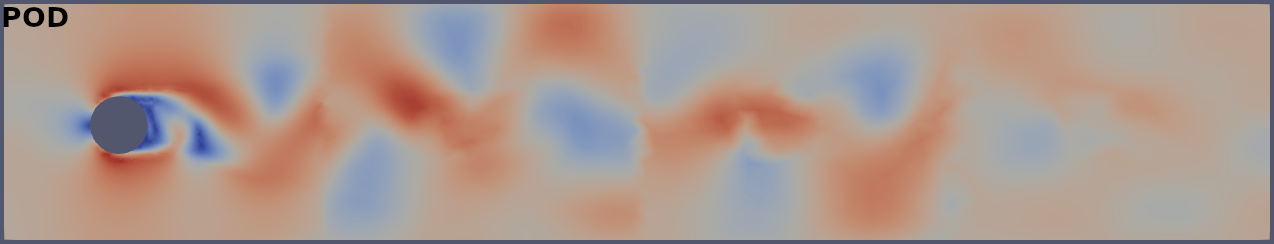}
    \includegraphics[width=0.45\textwidth]{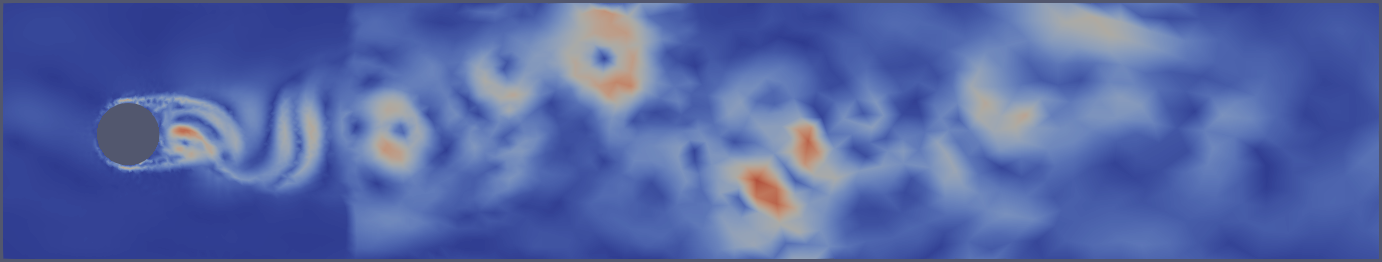}
    \includegraphics[width=0.45\textwidth]{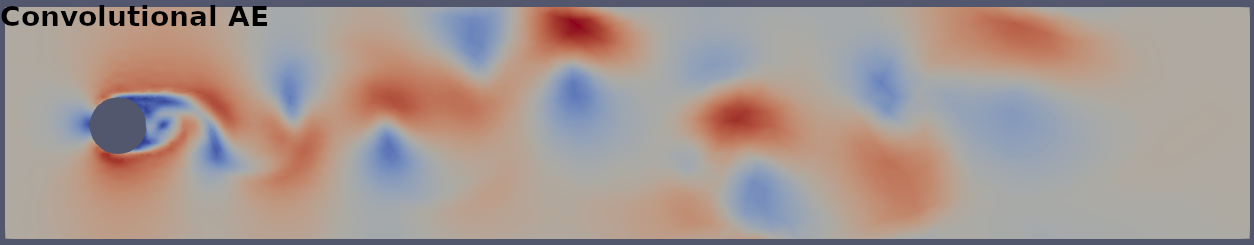}
    \includegraphics[width=0.45\textwidth]{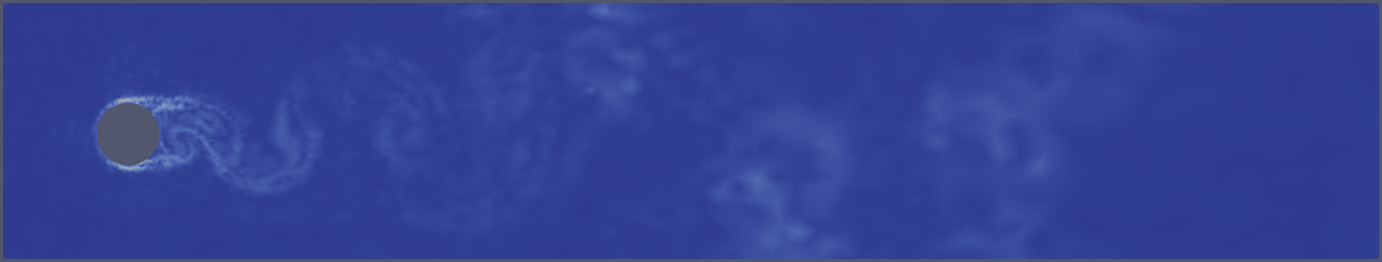}
    \includegraphics[width=0.45\textwidth]{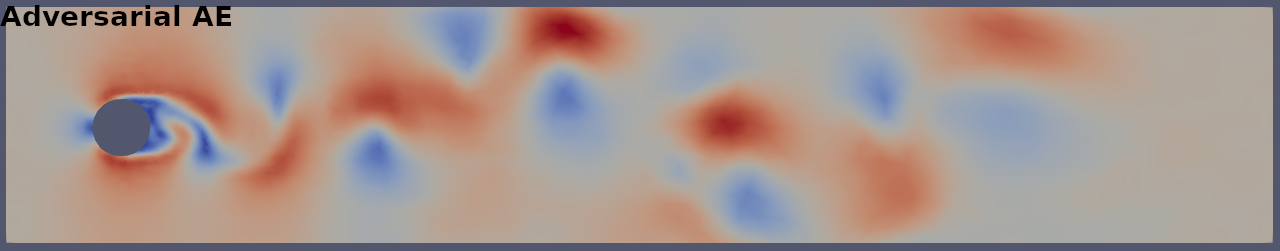}
    \includegraphics[width=0.45\textwidth]{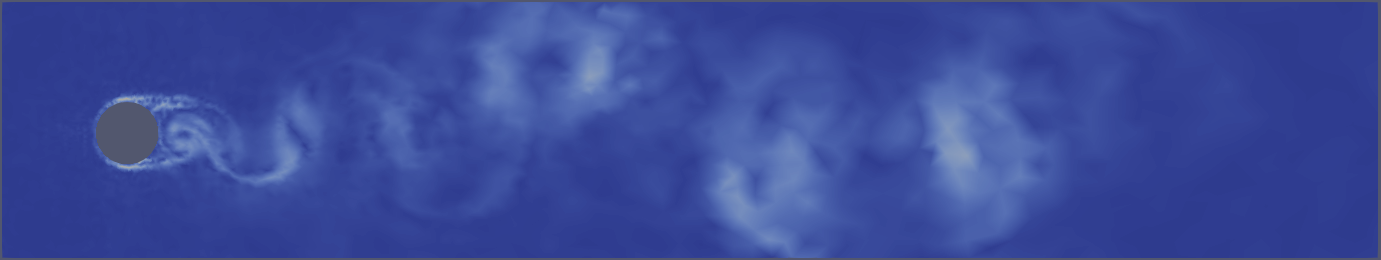}
    \includegraphics[width=0.45\textwidth]{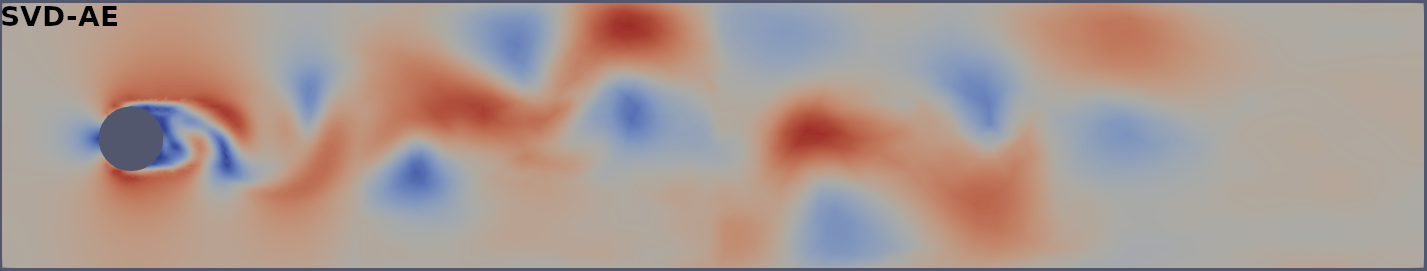}
    \includegraphics[width=0.45\textwidth]{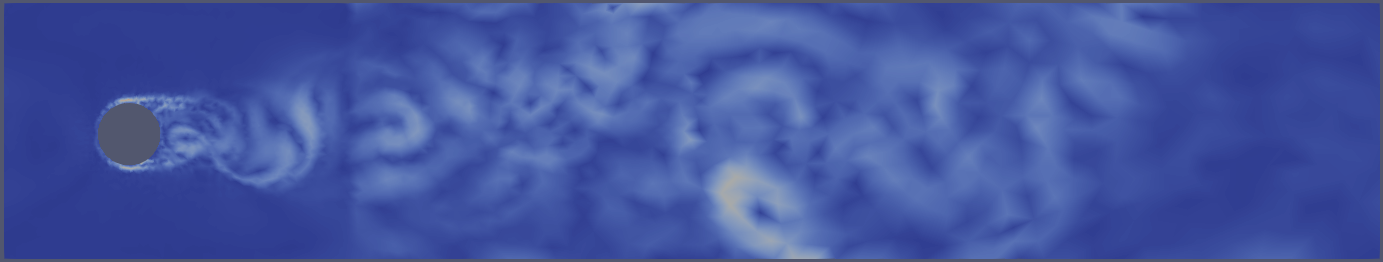}
    \includegraphics[width=0.45\textwidth]{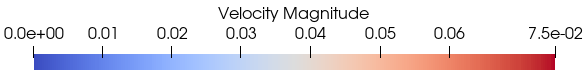}
    \includegraphics[width=0.45\textwidth]{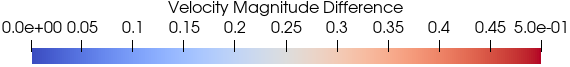}
    \caption{\label{fig:comparison_fpc}Velocity magnitude at a time of 50 seconds for the flow past a cylinder test case. From top to bottom (left): the original data; the reconstruction by POD; by the convolutional autoencoder; by the adversarial autoencoder; and by the SVD-Autoencoder. The pointwise errors are shown to the right of the reconstructions. Each of the reconstructions were made from 10 POD coefficients or latent variables. }
\end{figure*}

To test the methods, the solutions are compressed and reconstructed using Equation~\eqref{eq:recon_pod} for POD, Equation~\eqref{eq:recon_ae} for the convolutional and adversarial autoencoders, and Equation~\eqref{eq:recon_svdae} for the SVD autoencoder. The error in the reconstruction, Equation~\eqref{eq:recon_error_mse}, is calculated using the test dataset.  Figure~\ref{fig:comparison_fpc} shows the effect of the four compression methods (POD, CAE, AAE and SVD-AE) on a snapshot taken at  the 200th time level compared against the original snapshot. The pointwise errors in the velocity magnitude are shown on the right. It can be seen that all four methods (including POD) perform well in their reconstruction of flow past a cylinder. The pointwise errors indicate that, for this snapshot, the convolutional autoencoder gives the best results, followed by the adversarial autoencoder, the SVD-autoencoder and finally POD.  Table~\ref{table:compress_recon_mse_fpc} shows the mean of the square reconstruction errors calculated over the test dataset for the flow past a cylinder test case. 
As seen for the single snapshot in Figure~\ref{fig:comparison_fpc}, every compression method that involves an autoencoder outperforms POD. 

\begin{table}[htbp]
\caption{\label{table:compress_recon_mse_fpc}Reconstruction error averaged over the test dataset for flow past a cylinder using POD and several autoencoders. Each of the reconstructions were made from 10 POD coefficients or latent variables.}
\centering
\begin{tabular}{ cccc } 
\toprule
POD &  Convolutional AE & Adversarial AE & SVD-AE\\
\toprule
\num{111e-4} & \num{14.4e-4} & \num{62.9e-4} & \num{25.1e-4} \\
\bottomrule
\end{tabular}
\end{table}

\subsubsection{Multiphase flow in a pipe}
The domain is split into 10 subdomains (each spanning one tenth of the length ($x$) of the domain, but spanning the entire width ($y$) and height ($z$)), which are discretised with 60 by 20 by 20 structured grids. The velocity and volume fraction solutions from the unstructured mesh are interpolated onto these grids over 800 time levels each corresponding to \SI{0.01}{\second}. As before, the finite element basis functions are used to perform the interpolation. The dataset for this test case therefore has a total of 8000 samples (10 subdomains and 800 time levels). 

The four compression methods (POD, CAE, AAE, SVD-AE) are applied to the multiphase flow dataset. For POD, one column of the snapshots matrix consists of nodal values of the three velocity components (each scaled between -1 and 1) and the volume fractions (within the interval [0,1]). For the autoencoders, four channels are used and scaling is applied to the fields as usual. Initially 10 subdomains were used, however, as the autoencoders were found to have a relatively high error, a further 10 subdomains were created that were randomly located within the domain, making a total of 20 subdomains (and 16000 samples in the dataset). Having more subdomains provided more training data which probably led to the observed improvement in the results. The autoencoders were trained with 90\% of the data chosen at random from the dataset. For details of the hyperparameter optimsation and the networks used, see Tables~\ref{tab:app_hpo}, \ref{tab:app_optimal_ae} and~\ref{tab:architecture_DR} in the appendix.

Figure~\ref{fig:comparison_sf} shows how the autoencoders performed in reconstructing the pipe flow dataset. It is not surprising that they seem to perform less well than for the flow past a cylinder case, given the fact that the compression ratio was 80 for flow past a cylinder, whereas, for pipe flow, it was 9600. (For the former a 20 by 20 grid with two fields was compressed to 10 variables, whereas for the latter this a 60 by 20 by 20 grid with four fields was compressed to 10 variables.) Even at this compression ratio, all dimensionality reduction methods seemed able to reconstruct the slug in Figure~\ref{fig:comparison_sf} to some degree, with the convolutional AE doing this particularly well. For easier visualisation, Figure~\ref{fig:comparison_sf} shows just part of the domain, which includes a slug and also two boundaries between subdomains. The boundary at \SI{2}{m} can be identified by a slight kink that can be observed particularly well in the reconstructions of the AAE and the SVD-AE. This kink appears to the left of the slug, and highlights that for some models these boundaries induced additional inaccuracies. This issue could be addressed in future research by allowing the compressive methods to see the solutions of the neighbouring subdomains during compression, so that they can explicitly take this boundary into account.

\begin{figure*}[h!]
    \centering
    \includegraphics[width=0.49\textwidth]{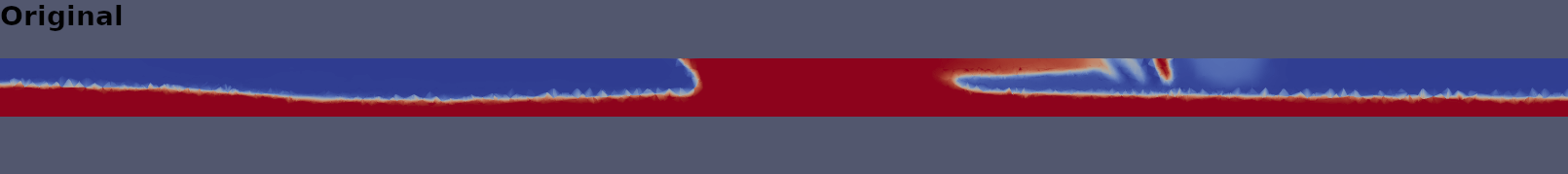}
    \includegraphics[width=0.49\textwidth]{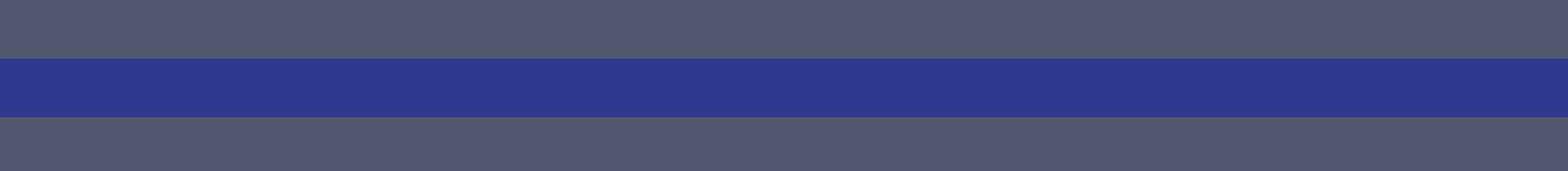}
    \includegraphics[width=0.49\textwidth]{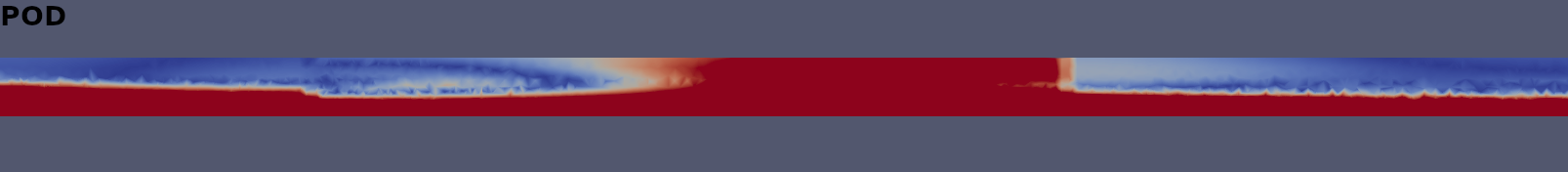}
    \includegraphics[width=0.49\textwidth]{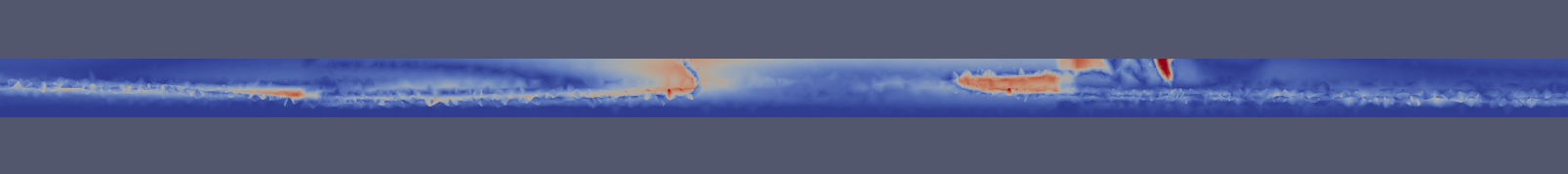}
    \includegraphics[width=0.49\textwidth]{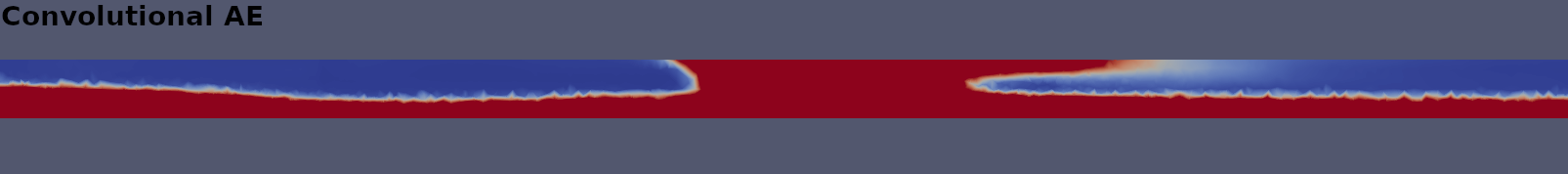}
    \includegraphics[width=0.49\textwidth]{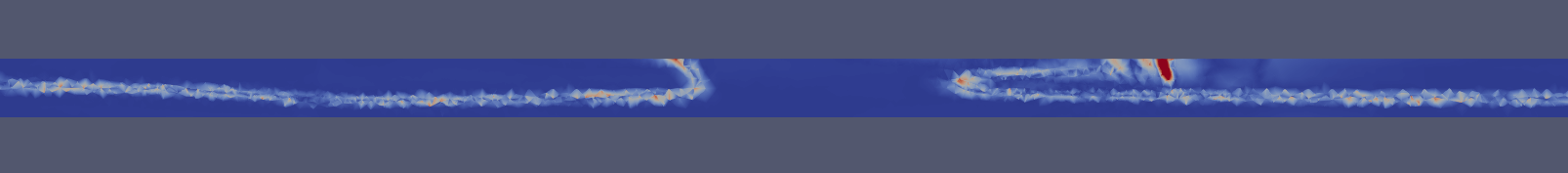}
    \includegraphics[width=0.49\textwidth]{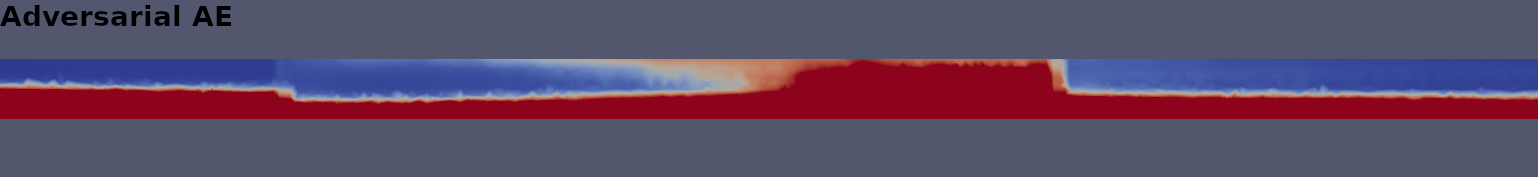}
    \includegraphics[width=0.49\textwidth]{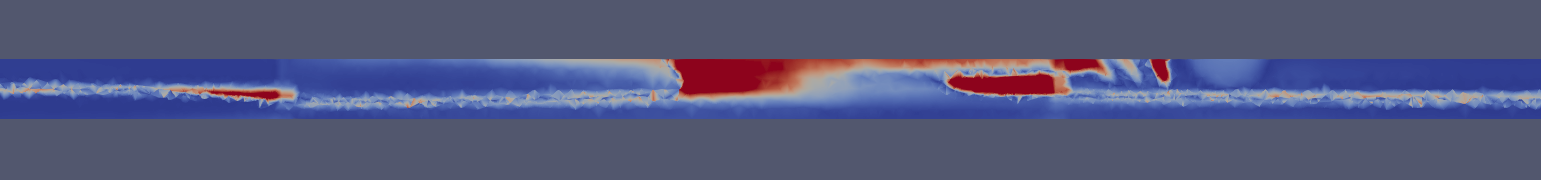}
    \includegraphics[width=0.49\textwidth]{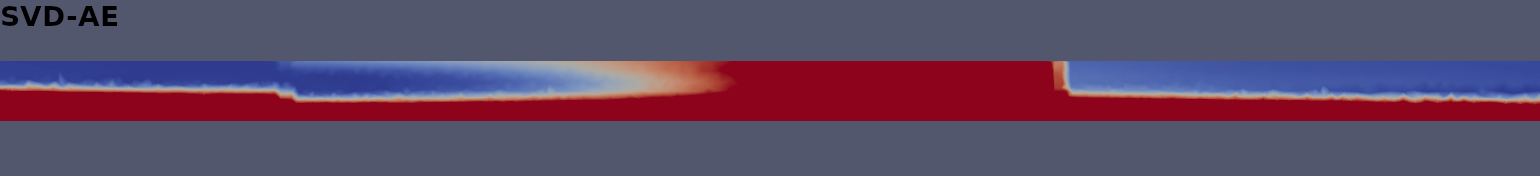}
    \includegraphics[width=0.49\textwidth]{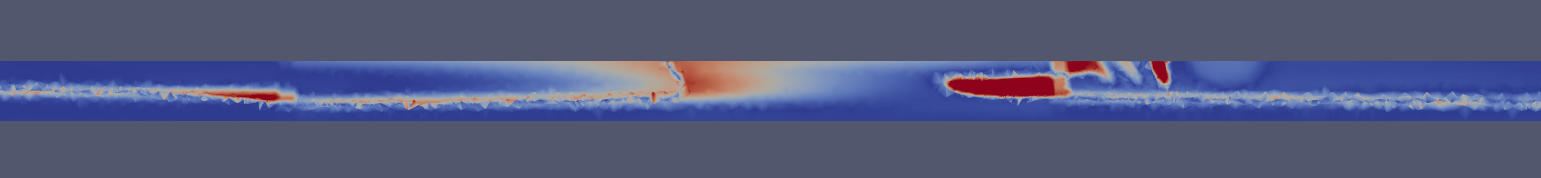}
    \includegraphics[width=0.49\textwidth]{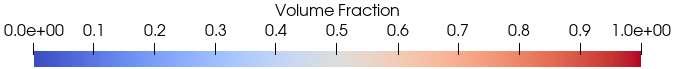}
    \includegraphics[width=0.49\textwidth]{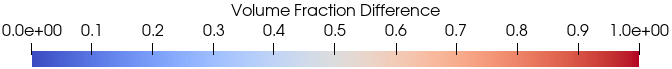}
    \caption{The volume fractions taken at a time of \SI{1.73}{s}, spanning the domain between \SI{1.59}{m} and \SI{3.67}{m}. A cross-section along the length through the centre of the pipe is shown. From top to bottom (left), the snapshots are from the original pipe flow data; the data reconstructed by POD; by the convolutional autoencoder; by the adversarial autoencoder; and by the SVD-Autoencoder. The pointwise errors are plotted to the right of the reconstructions. Each of the reconstructions were made from 10 POD coefficients or latent variables.}
    \label{fig:comparison_sf}
\end{figure*}

Table~\ref{table:compress_recon_mse} shows the reconstruction error over the test data for the dimensionality reduction methods. Here, Equation~\eqref{eq:recon_error_mse} was used, where vectors $\bm{u}^k$ and $(\bm{u}^{\text{recon}})^k$ consists of the scaled velocities and volume fractions. Once again, the convolutional autoencoder has the lowest errors.

\begin{table}[htbp]
\caption{\label{table:compress_recon_mse}Reconstruction error for POD and the autoencoders over the test dataset. Each of the reconstructions were made from 10 POD coefficients or latent variables.}
\centering
\begin{tabular}{ cccc } 
\toprule
 POD & Convolutional AE & Adversarial AE & SVD-AE  \\
\toprule
 \num{21.7e-4} & \num{4.70e-4} & \num{20.2e-4} & \num{30.8e-4}  \\
\bottomrule
\end{tabular}
\end{table}

\subsection{Prediction for multiphase flow in a pipe}
As the convolutional autoencoder performed better than the other networks for dimensionality reduction, we go on to combine this with a predictive adversarial network within a domain decomposition framework to form a reduced-order model (AI-DDNIROM). %
A schematic diagram of how the networks are combined can be seen in Figure~\ref{fig:schematic_workflow}. 

\subsubsection{Training and predicting with the original domain}

\begin{figure*}[h!]
    \centering
    \includegraphics[width=0.49\textwidth]{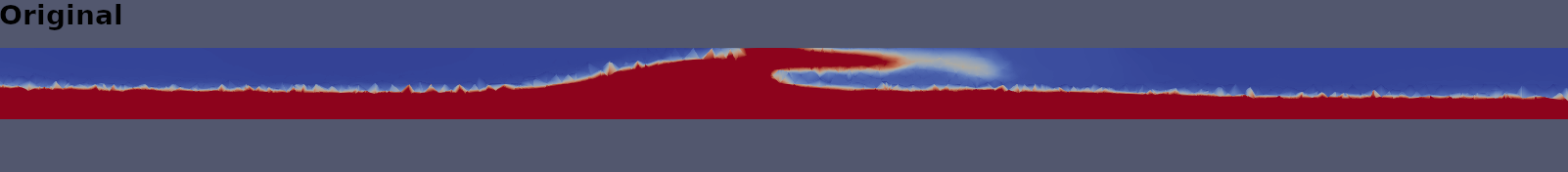}
    \includegraphics[width=0.49\textwidth]{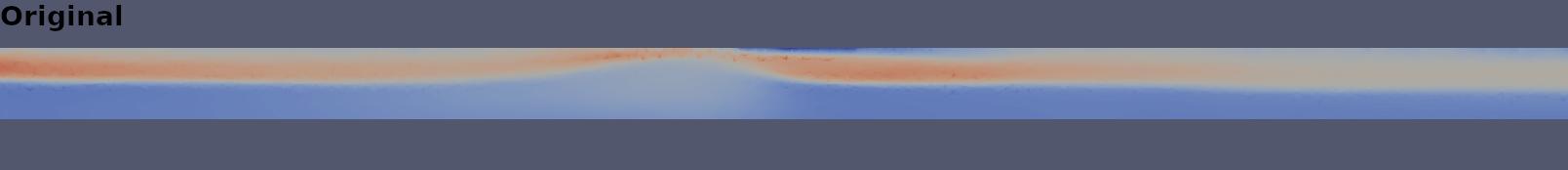}
    \includegraphics[width=0.49\textwidth]{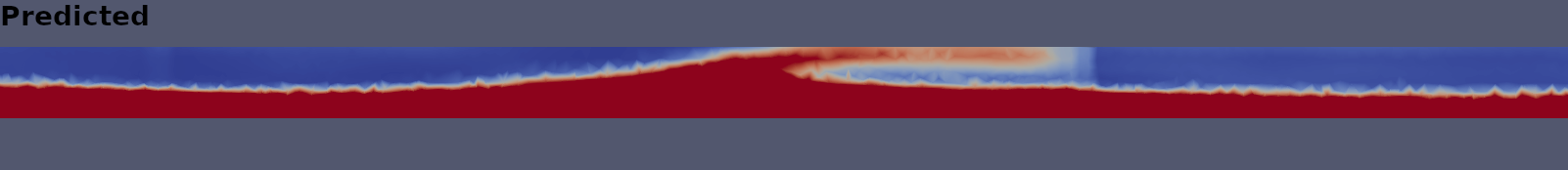}
    \includegraphics[width=0.49\textwidth]{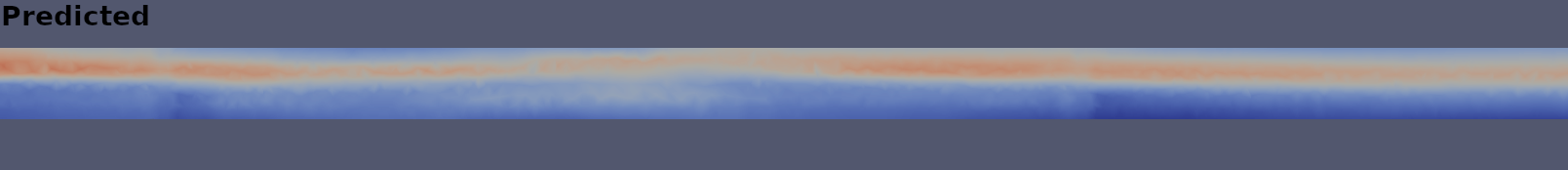}
    \includegraphics[width=0.49\textwidth]{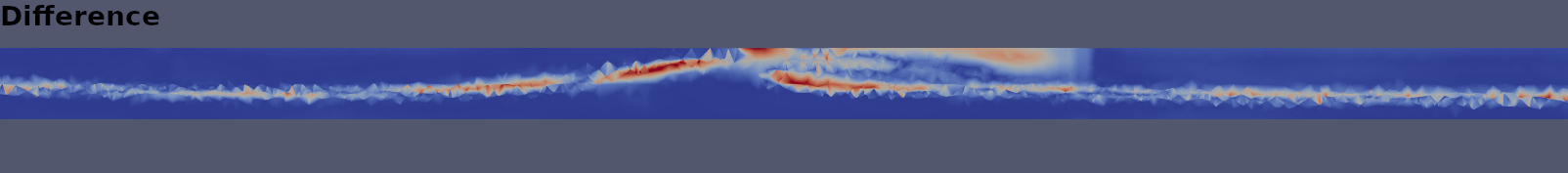}
    \includegraphics[width=0.49\textwidth]{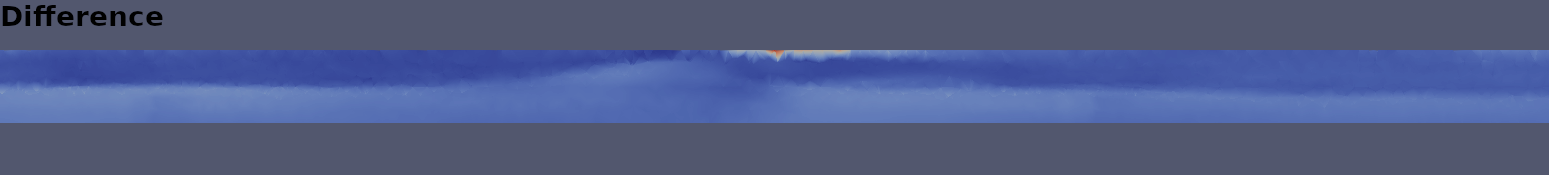}
    \includegraphics[width=0.49\textwidth]{legend_sf_volfrac.png}
    \includegraphics[width=0.49\textwidth]{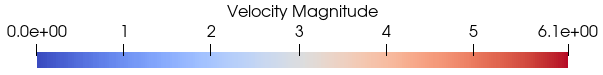}
    \caption{A snapshot of the volume fractions (left) and the velocity fields (right) at $t=$\SI{8.96}{s}, spanning the domain between \SI{2.86}{m} and \SI{4.51}{m} and sliced exactly through the middle.The top plots show the original CFD, the middle plots show the predictions from the AI-DDNIROM and the bottom plots show the pointwise error between the CFD and the AI-DDNIROM.}
    \label{fig:prediction_original_domain_vof}
\end{figure*}

For the prediction, the HFM produced 1400 solutions over \SI{14}{seconds} of real time. The training and validation data was taken from time levels 1 to 799, and the test data from time levels 800 to 1400. Hyperparameter optimisation was performed and the results of this can be found in Tables~\ref{tab:app_optimal_pan} and~\ref{tab:app_optimal_pan_arch} of the appendix. AS part of this process, it was found that the best time step for the NIROM was \SI{0.06}{\second}, i.e.~6 times as large as the time interval between the HFM solutions. The MSE achieved on the validation data was \num{15.6e-4} and on the test data was \num{103e-4}. Figure~\ref{fig:prediction_original_domain_vof} compares the predictions of volume fraction with those of the HFM and shows the pointwise error for two snapshots in the test data (unseen by the model). The agreement between the predictive adversarial network and the HFM is very good. 

\subsubsection{Extending the domain and associated boundary conditions}
\begin{figure*}[htbp]
\centering
\includegraphics[width=0.95\textwidth]{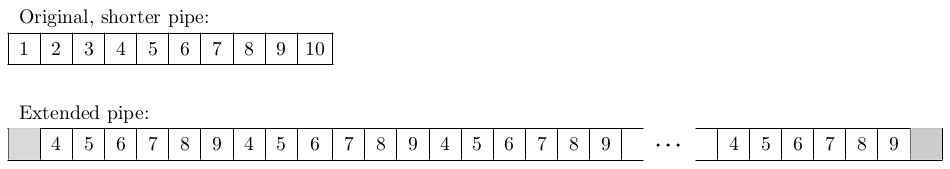}
\caption{\label{fig:pipe_and_longer_pipe}Above: the shorter, original pipe used in generating the snapshots with subdomain numbering. Below: the extended pipe with initial conditions taken from the indicated subdomains of the shorter pipe. The grey subdomains at either end take their initial conditions from the boundary conditions.}
\end{figure*}

Having trained an AI-DDNIROM in the previous section with snapshots from the \SI{10}{\metre} long pipe and made predictions for that pipe, in this section we use the method described in Section~\ref{sec:extension} to predict the flow evolution and volume fractions along a pipe of length \SI{98}{\metre} based on training data from the \SI{10}{\metre} pipe. The extended pipe is split into 98 subdomamins, for which the initial conditions come from the simulation of the \SI{10}{\metre} pipe taken at \SI{7.2}{\second} (time level 720). This is in order to start simulating from a state that is well developed. The first subdomain of the \SI{98}{\metre} pipe takes initial conditions from the third subdomain of the \SI{10}{\metre} pipe; the second to seventh subdomains of the \SI{98}{\metre} pipe take the values from the fourth to the ninth subdomains of the \SI{10}{\metre} pipe. This is repeated 15 more times, and the final subdomain of the \SI{98}{\metre} pipe takes the tenth and final subdomain of the \SI{10}{\metre} pipe, see Figure~\ref{fig:pipe_and_longer_pipe}. The first, second  and tenth subdomains of the \SI{10}{\metre} pipe were not used, to avoid introducing any spurious effects from the boundaries.

Velocity and volume fractions are specified throughout time in the first and last (98th) subdomains which act in a manner similar to boundary conditions. There is no high-fidelity model for the \SI{98}{m} pipe from which to take boundary conditions, and, as the time over which predictions are made exceeds the time over which snapshots were collected from the high-fidelity model of the \SI{10}{m} pipe, boundary conditions must be generated somehow. Three methods of producing boundary conditions are reported (as described in Section~\ref{sec:extension}): \begin{enumerate}[(i)]
    \item Cycling through slug formation: a slug is found in the shorter pipe, and the velocity and volume fraction fields associated with the advection of this slug through the subdomain are repeated as required. The particular subdomain of the shorter pipe was the third (between \SI{2}{m} and \SI{3}{m}), between time levels 750 to 804. So, the boundary condition for the left-most end of the extended pipe can be written as
    \begin{eqnarray}
    \bm{\alpha}^{\text{ext}}_1(t_k) &=& \bm{\alpha}_3(\tilde{t}_k)  \quad \forall t_k\geqslant 0\,,\\
    \bm{u}^{\text{ext}}_1(t_k) &=& \bm{u}_3(\tilde{t}_k)  \quad \forall t_k\geqslant 0\,, 
    \end{eqnarray}
    where $t_k = k \Delta t$ for time level~$k$, ($k=0,1,\ldots$) and a time step of $\Delta t$, and 
    \begin{equation}
        \tilde{t}_k = \left( \frac{t_k}{\Delta t}\Mod{54}+750 \right)\Delta t\,.
    \end{equation}
    where $a\Mod{n}$ gives the non-negative remainder when $n$ has been subtracted from $a$ as many times as possible. 
    For this example, the time step of the reduced-order model is \SI{0.06}{s}. The slug appears in this subdomain shortly after the selected time window as a relatively thin instability, of the order of magnitude of \SI{10}{cm} in length, and develops in width as it advects through the domain. 
    \item Perturbed instability: at the 798th time level an instability occurs in the third subdomain of the shorter pipe. The volume fraction field associated with this is perturbed spatially by Gaussian noise, the velocity field is left unperturbed, and both are used as boundary conditions in the first subdomain of the extended pipe. In the following, $\bm{\alpha}^{\text{ext}}_1$ is the volume fraction in the first subdomain of the extended pipe, $\bm{\alpha}_3$ is the volume fraction in the third subdomain of the shorter pipe
    \begin{alignat}{3}
        & \bm{\alpha}^{\text{ext}}_1(t_k) =&&\ \bm{\alpha}_3(\tilde{t}_k) + r  \qquad && \forall t_k\geqslant 0\,,\\
        & \bm{u}^{\text{ext}}_1(t_k) =&&\ \bm{u}_3(\tilde{t}_k) \qquad  && \forall t_k\geqslant 0 \,,
    \end{alignat}
    where $\tilde{t}_k = $ \SI{7.98}{\second} and $r$ is a random spatial perturbation.  
    \item Original boundaries repeated: velocity and volume fraction solutions from the third subdomain of the shorter pipe are used as the boundary conditions for the first subdomain in the extended pipe and repeated for as long as required. The solution fields from time \SI{1}{\second} to \SI{8}{\second} are used, as this corresponds to times where the air had passed through the entire length of the shorter pipe. Therefore, for times in $[0,7)$ seconds of the extended pipe, times in $[1,8)$ seconds of the shorter pipe are used; for times in [7,14) seconds of the extended pipe, times in [1,8) seconds of the shorter pipe are used; etc. So, the boundary condition for the left-most end of the extended pipe can be written as
    \begin{eqnarray}
    \bm{\alpha}^{\text{ext}}_1(t_k) &=& \bm{\alpha}_3(\tilde{t}_k)  \quad \forall t_k\geqslant 0\,,\\
    \bm{u}^{\text{ext}}_1(t_k) &=& \bm{u}_3(\tilde{t}_k)  \quad \forall t_k\geqslant 0\,, 
    \end{eqnarray}
    where $t_k = k \Delta t$ for time level~$k$, ($k=0,1,\ldots$) and a time step of $\Delta t$, and 
    \begin{equation}
        \tilde{t}_k = \left( \frac{t_k}{\Delta t}\Mod{700}+100 \right)\Delta t\,.
    \end{equation}
\end{enumerate}
In all cases, the boundary condition for the final subdomain is based on a snapshot from subdomain~2 at \SI{7.5}{\second} in the \SI{10}{m} pipe when the flow was almost steady with the lower half of the pipe occupied by water and the upper half occupied by air.

\begin{figure*}[htbp]
\begin{subfigure}{.33\textwidth}
  \centering
  \includegraphics[width=1\linewidth]{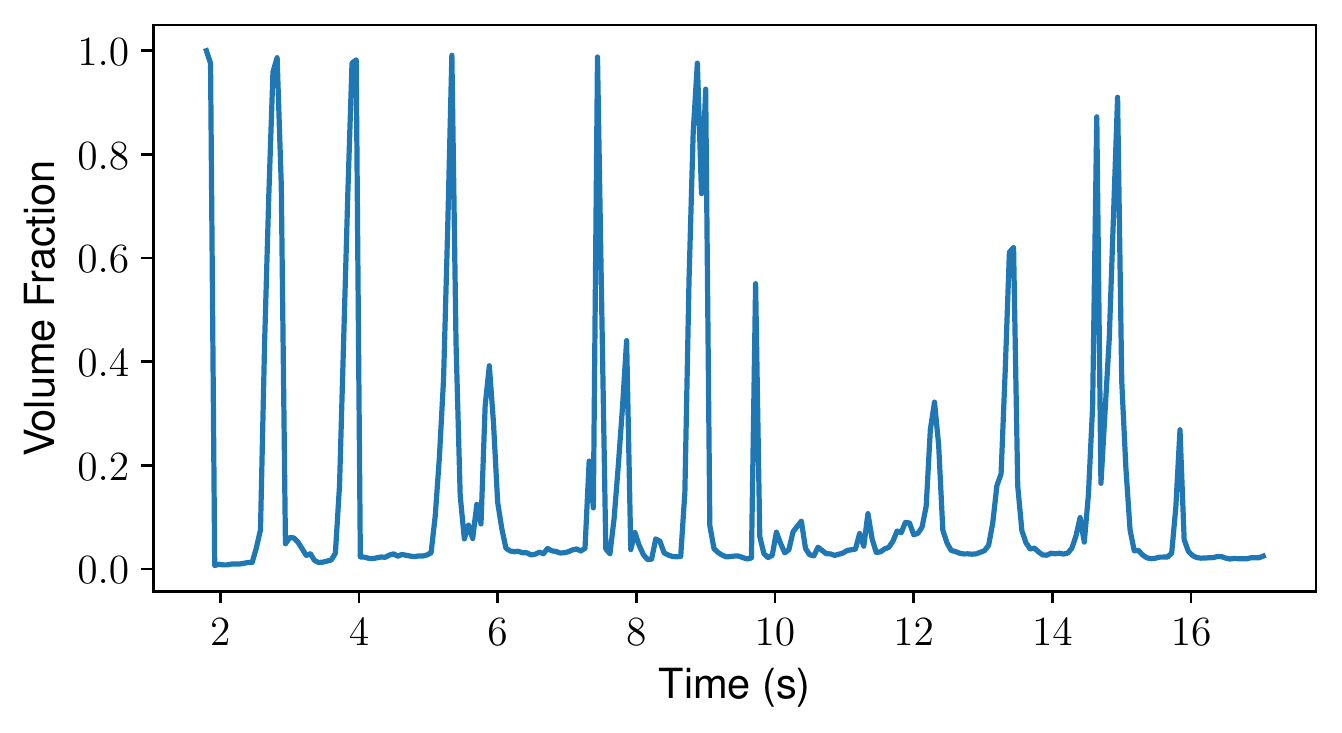}
  \caption{Original simulation}
  \label{fig:comparison_mass_frac_a}
\end{subfigure}%
\begin{subfigure}{.33\textwidth}
  \centering
  \includegraphics[width=1\linewidth]{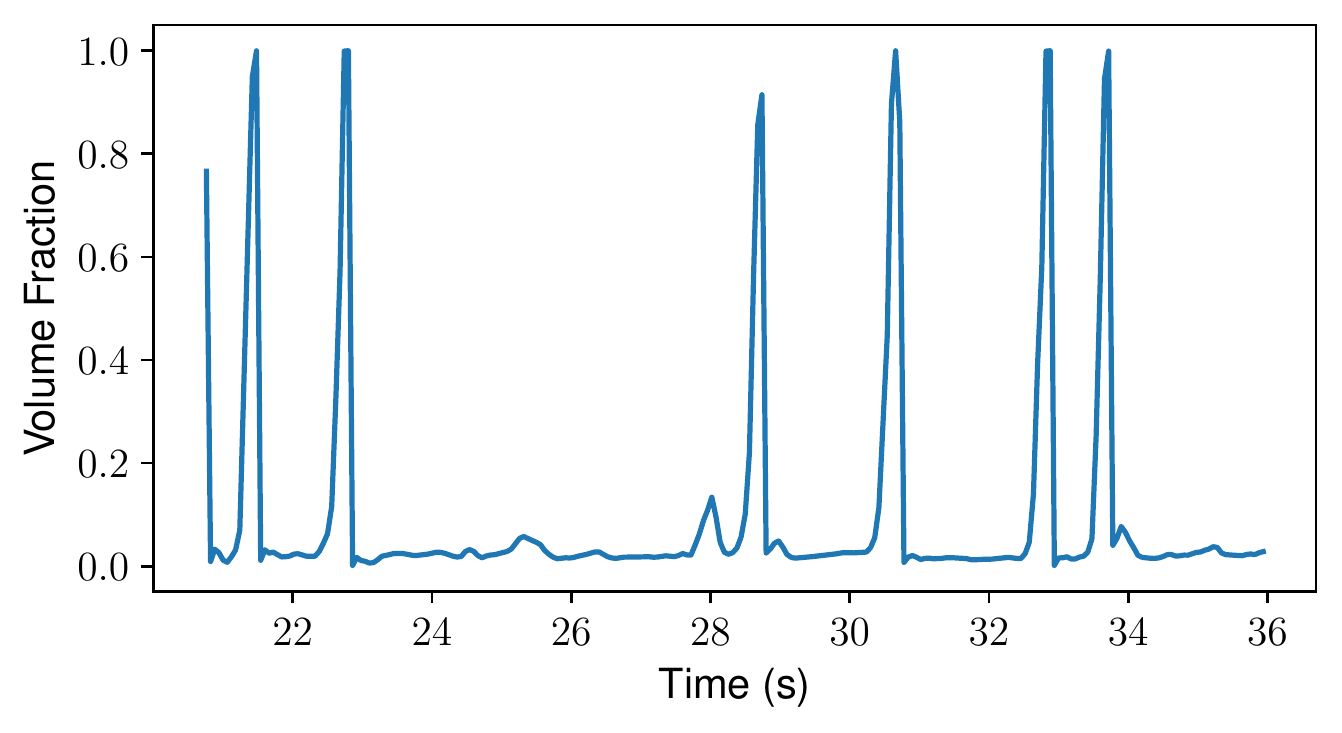}
  \caption{Cycling through slug formation}
  \label{fig:comparison_mass_frac_b}
\end{subfigure}
\begin{subfigure}{.33\textwidth}
  \centering
  \includegraphics[width=1\linewidth]{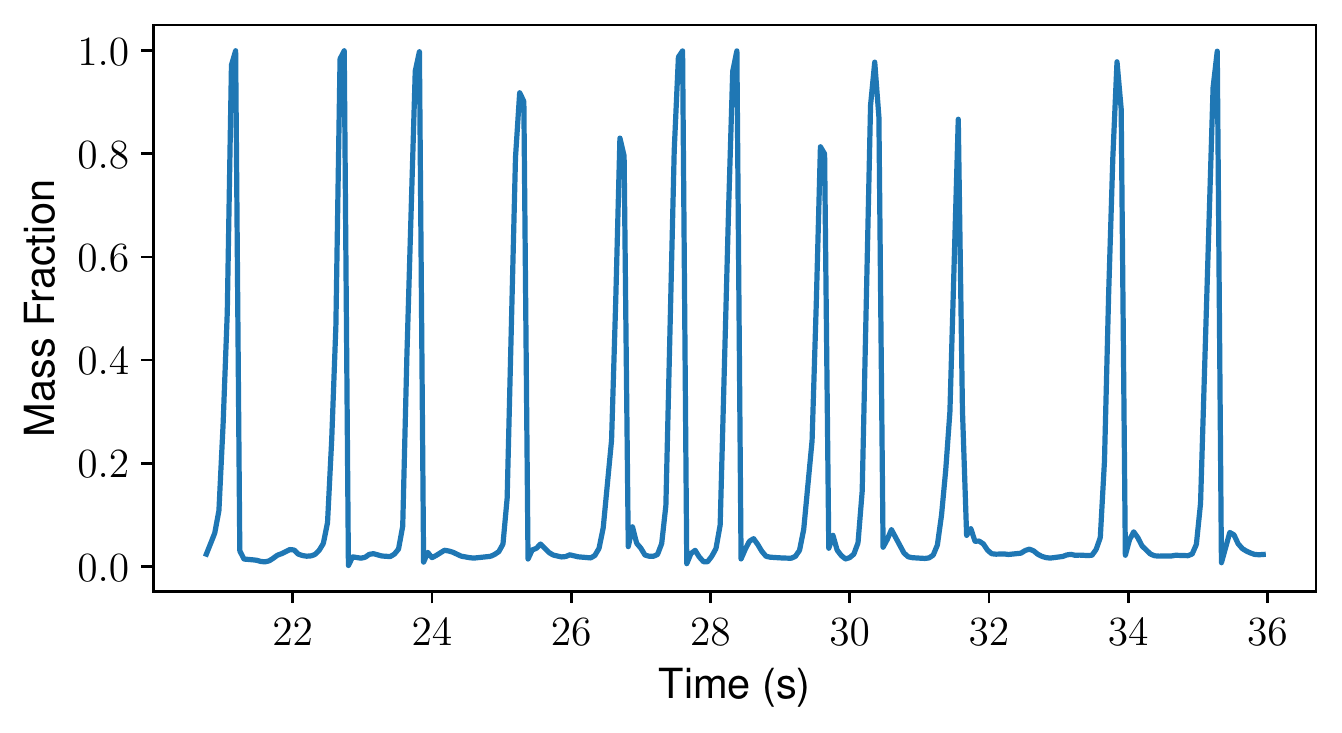}
  \caption{Perturbed instability}
  \label{fig:comparison_mass_frac_c}
\end{subfigure}
\begin{subfigure}{.33\textwidth}
  \centering
  \includegraphics[width=1\linewidth]{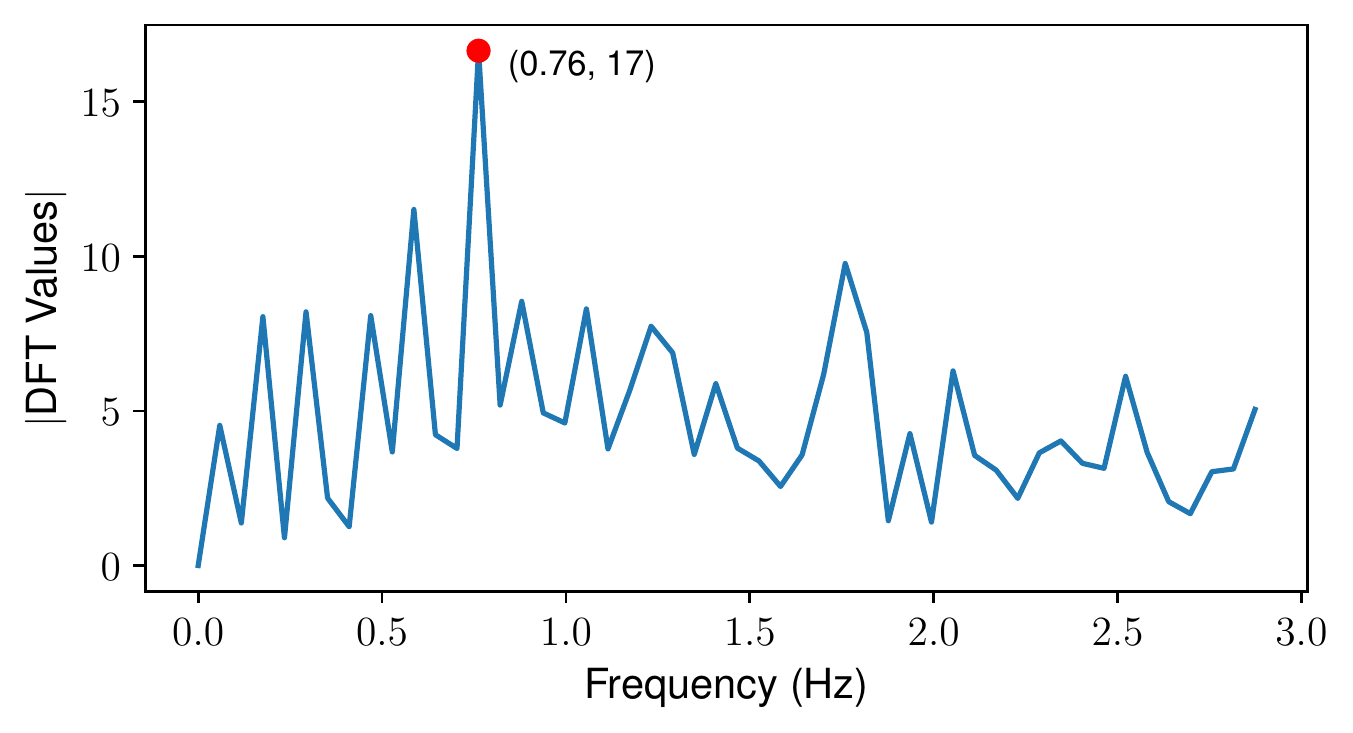}
  \caption{Original simulation}
  \label{fig:comparison_mass_frac_d}
\end{subfigure}%
\begin{subfigure}{.33\textwidth}
  \centering
  \includegraphics[width=1\linewidth]{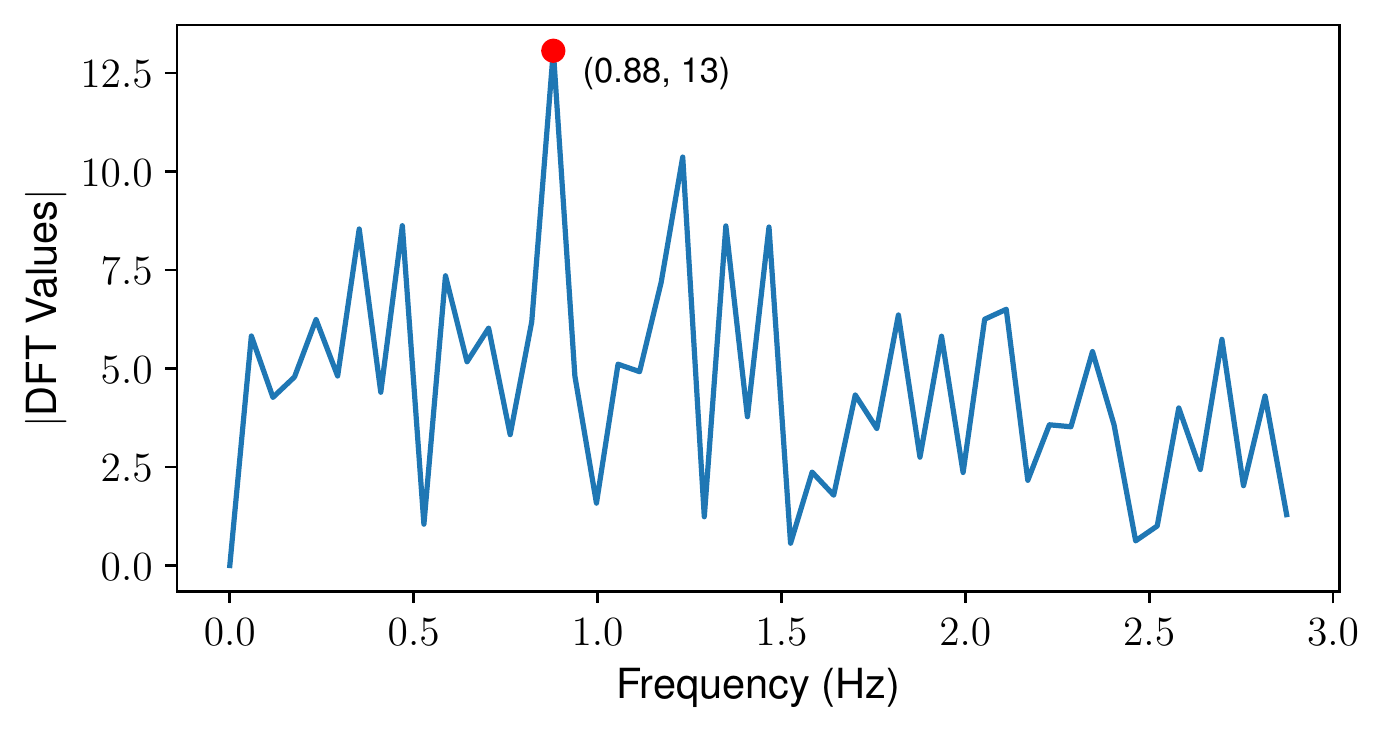}
  \caption{Cycling through slug formation}
  \label{fig:comparison_mass_frac_e}
\end{subfigure}
\begin{subfigure}{.33\textwidth}
  \centering
  \includegraphics[width=1\linewidth]{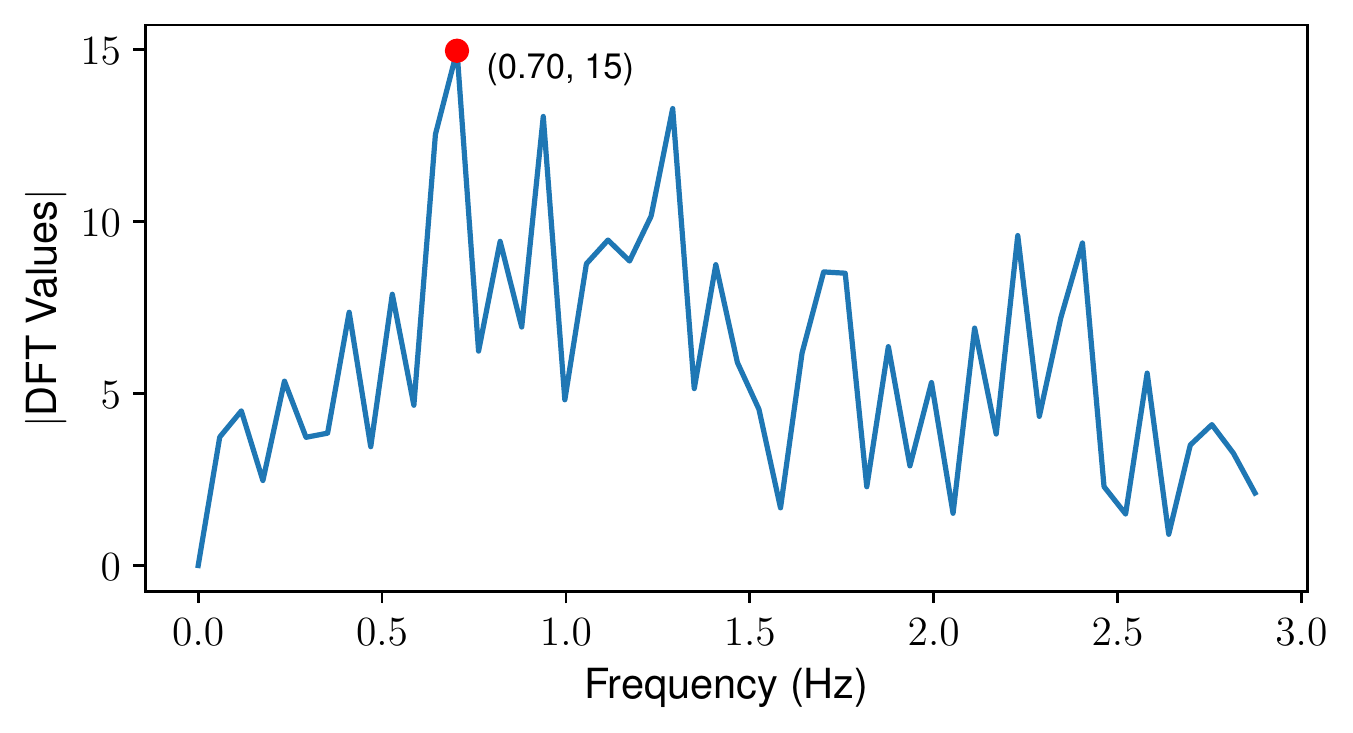}
  \caption{Perturbed instability}
  \label{fig:comparison_mass_frac_f}
\end{subfigure}
\caption{(a, b, c) Volume fractions plotted against time at (a) \SI{6.5}{m} for the original results and at (b,c) \SI{64.5}{m} for the reduced models. In each case, the data was collected at a height of 0.1$r$ = \SI{0.0039}{\metre} above the centreline of the pipe. (d, e, f) Discrete Fourier Transform (DFT) applied to the data presented in subfigures (a, b, c). }
\label{fig:comparison_mass_frac}
\end{figure*}

Various statistics are presented in this section in an aim to assess whether the AI-DDNIROM approach produces realistic results. If the expected advantage of the adversarial training strategy to produce a model that does not extrapolate beyond the seen training data holds true, then the predictive model could be expected to not diverge significantly from the original simulation. Figures~\ref{fig:comparison_mass_frac_a},  \ref{fig:comparison_mass_frac_b}, and~\ref{fig:comparison_mass_frac_c} show how the liquid volume fraction field varies over time in the original simulation and for the reduced models using two of the tested boundary conditions (cycling through slug formation and perturbed instability). The results obtained when repeating the original boundary conditions were similar to the original simulation and are not shown here. The time interval for the two reduced-order models corresponds to the instabilities having passed through two thirds of the pipe.  
Time series data was collected at values of $x=\SI{6.5}{m}$ (for the original simulation) and $x=\SI{64.5}{m}$ (for the two reduced models), and at a height of \SI{0.0039}{\metre} (a tenth of the pipe radius) above the centreline of the pipe. To analyse the frequency spectra, a discrete Fourier transform was then applied to the data. Figures~\ref{fig:comparison_mass_frac_d}, \ref{fig:comparison_mass_frac_e}, and~\ref{fig:comparison_mass_frac_f} show that the slug characteristic frequency spectra for the predictions are similar to that of the original simulation. In particular the main peak has a similar value in all three simulations (original simulation: \SI{0.76}{\hertz}; reduced model which cycles through slug formation: \SI{0.7}{\hertz}; reduced model with the perturbed instability \SI{0.88}{\hertz}). This suggests that simulations from the AI-DDNIROMs based on either of these boundary conditions are able to behave in a realistic way. In fact, the frequency of the main peak could be interpreted as the pseudo-slug frequency. Technically slugs are only defined as such when they span the full vertical extent of the pipe. On the other hand, pseudo-slugs~\cite{Fan2020} or proto-slugs~\cite{friedemann2019gas} are precursors to slugs, which do not necessarily reach the full height of the pipe.

\begin{figure*}[htpb]
\centering
\begin{subfigure}{.33\textwidth}
  \centering
  \includegraphics[width=1\linewidth]{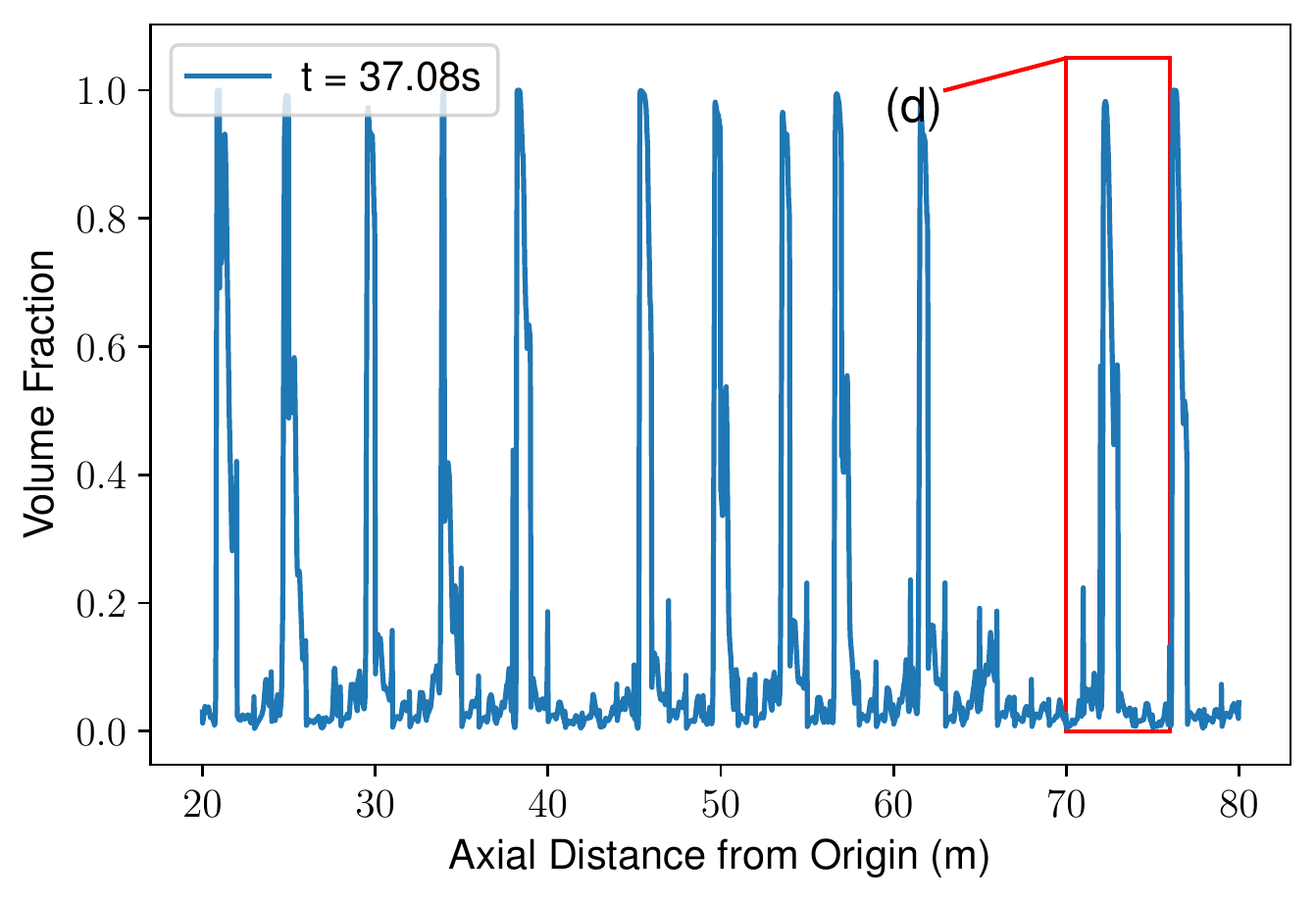}
  \caption{Cycling through slug formation}
  \label{fig:mass_frac_snapshots_a}
\end{subfigure}%
\begin{subfigure}{.33\textwidth}
  \centering
  \includegraphics[width=1\linewidth]{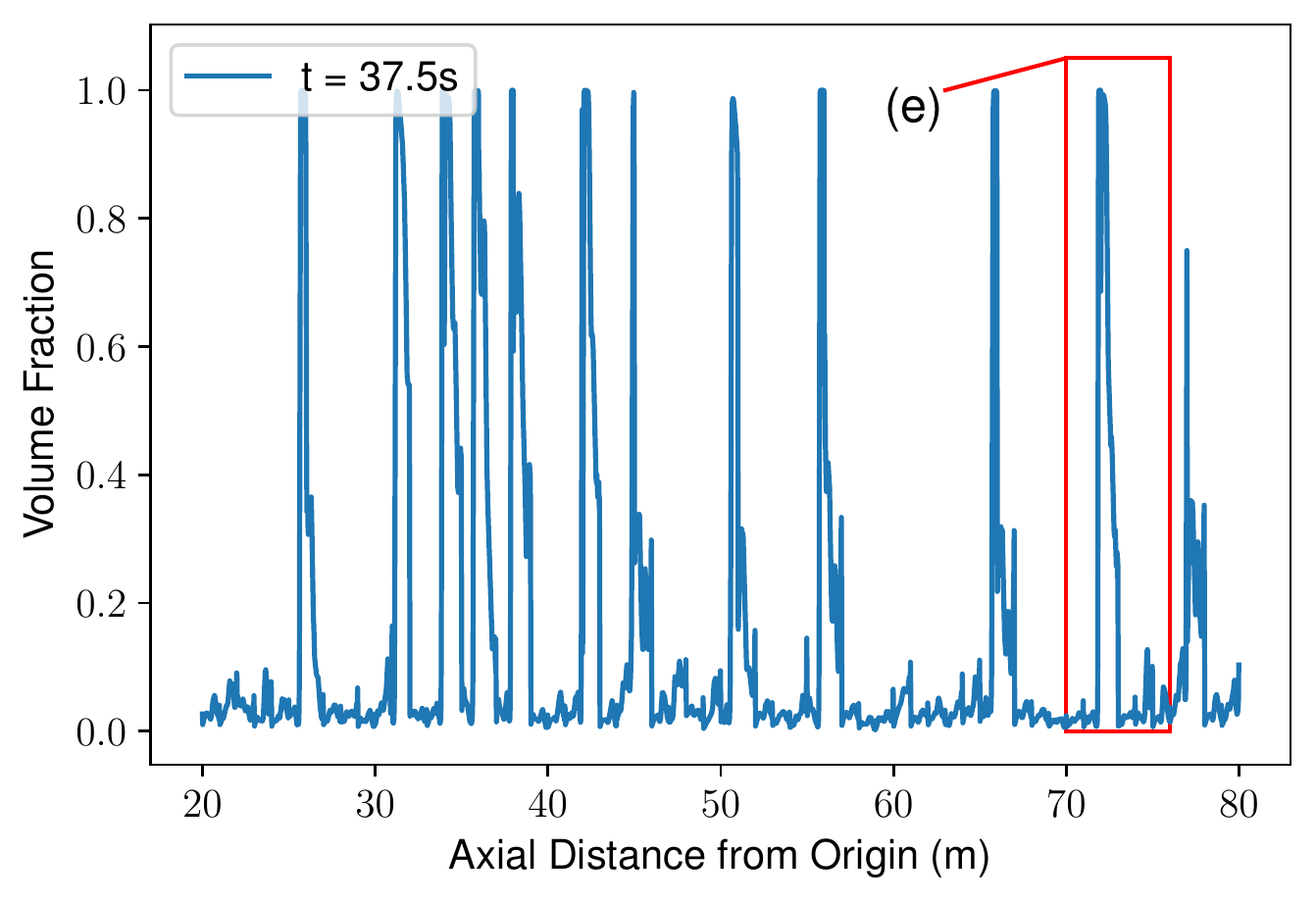}
  \caption{Perturbed instability}
  \label{fig:mass_frac_snapshots_b}
\end{subfigure}%
\newline
\begin{subfigure}{.33\textwidth}
  \centering
  \includegraphics[width=1\linewidth]{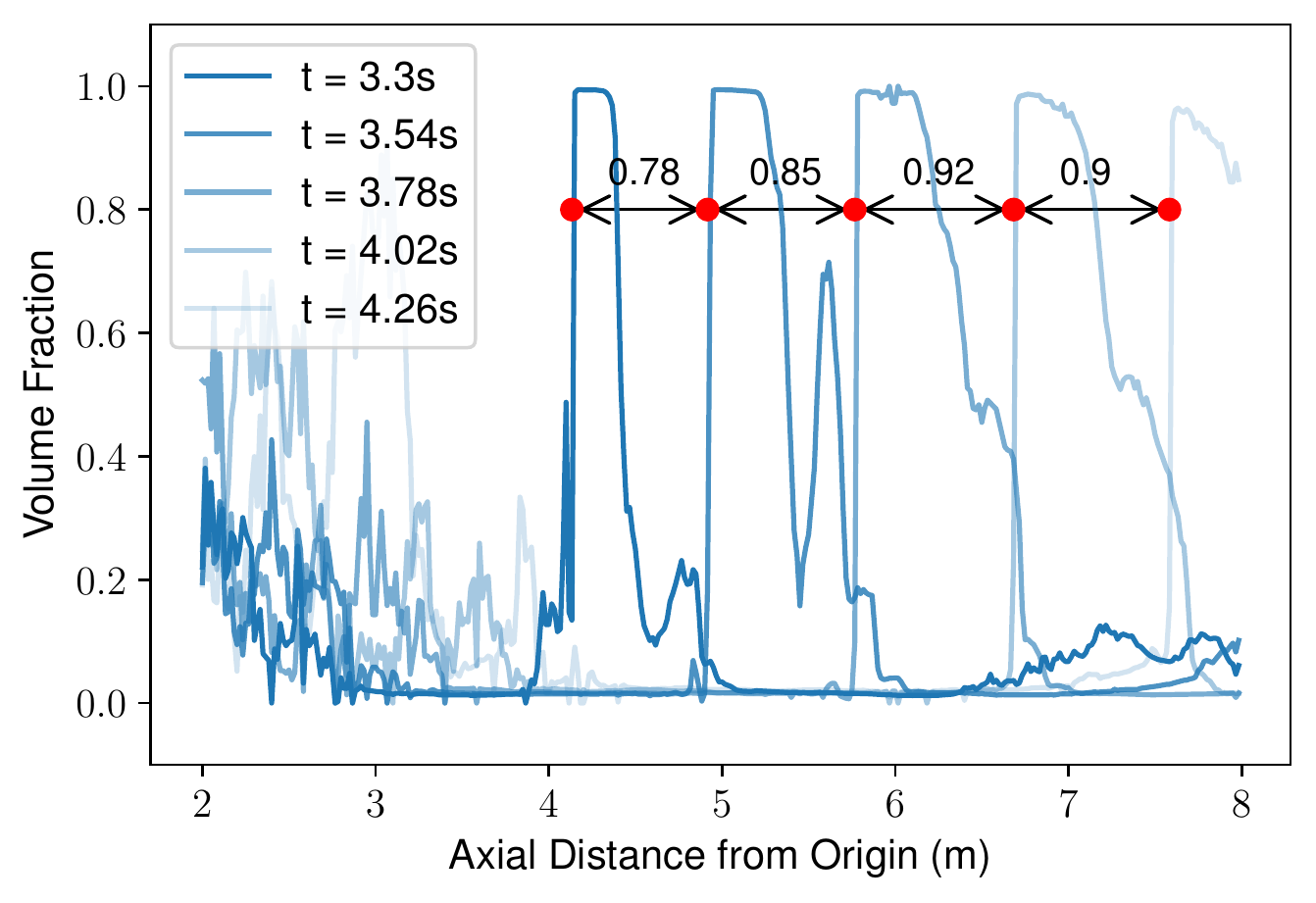}
  \caption{Original simulation}
  \label{fig:mass_frac_snapshots_c}
\end{subfigure}%
\begin{subfigure}{.33\textwidth}
  \centering
  \includegraphics[width=1\linewidth]{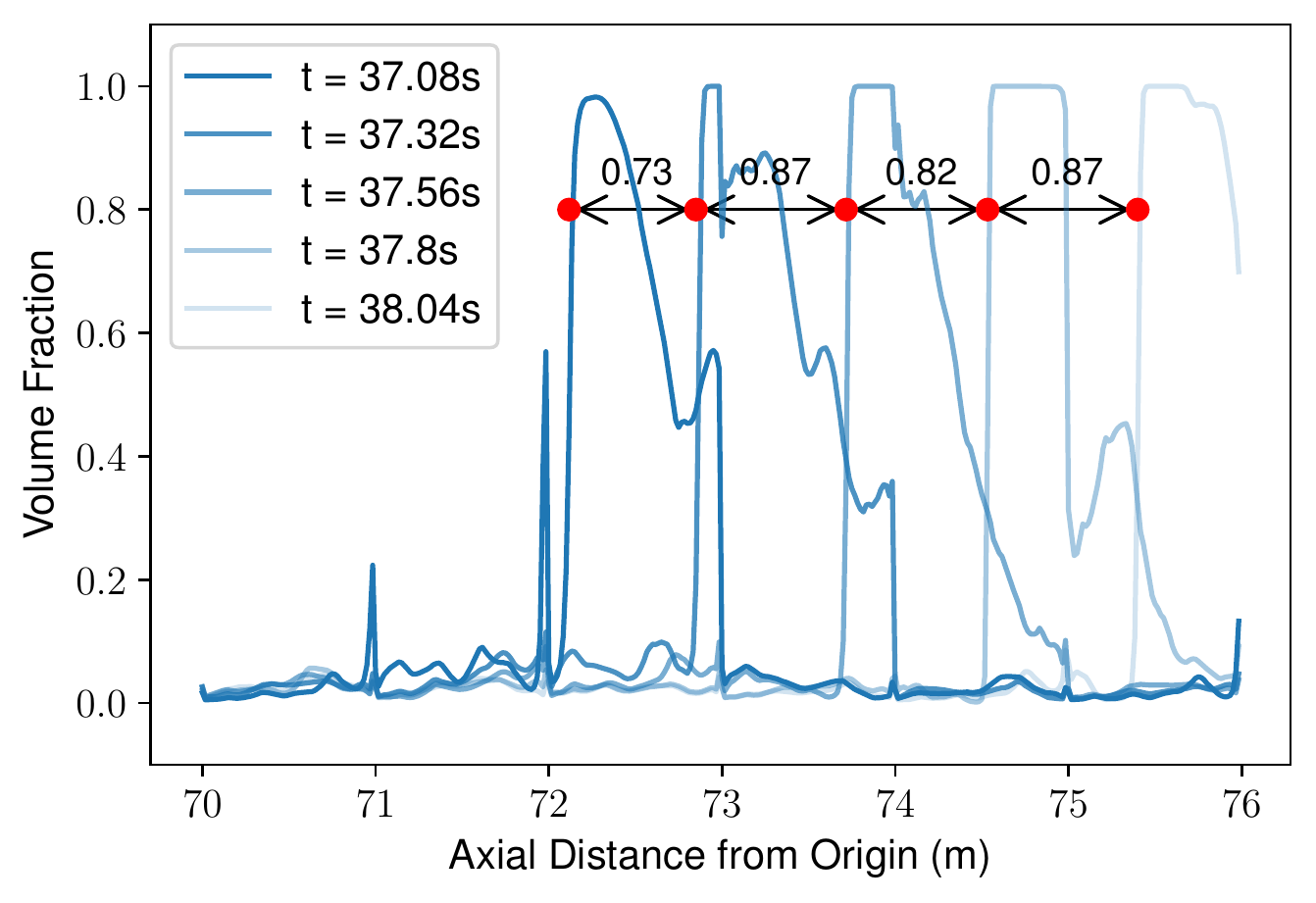}
  \caption{Cycling through slug formation}
  \label{fig:mass_frac_snapshots_d}
\end{subfigure}
\begin{subfigure}{.33\textwidth}
  \centering
  \includegraphics[width=1\linewidth]{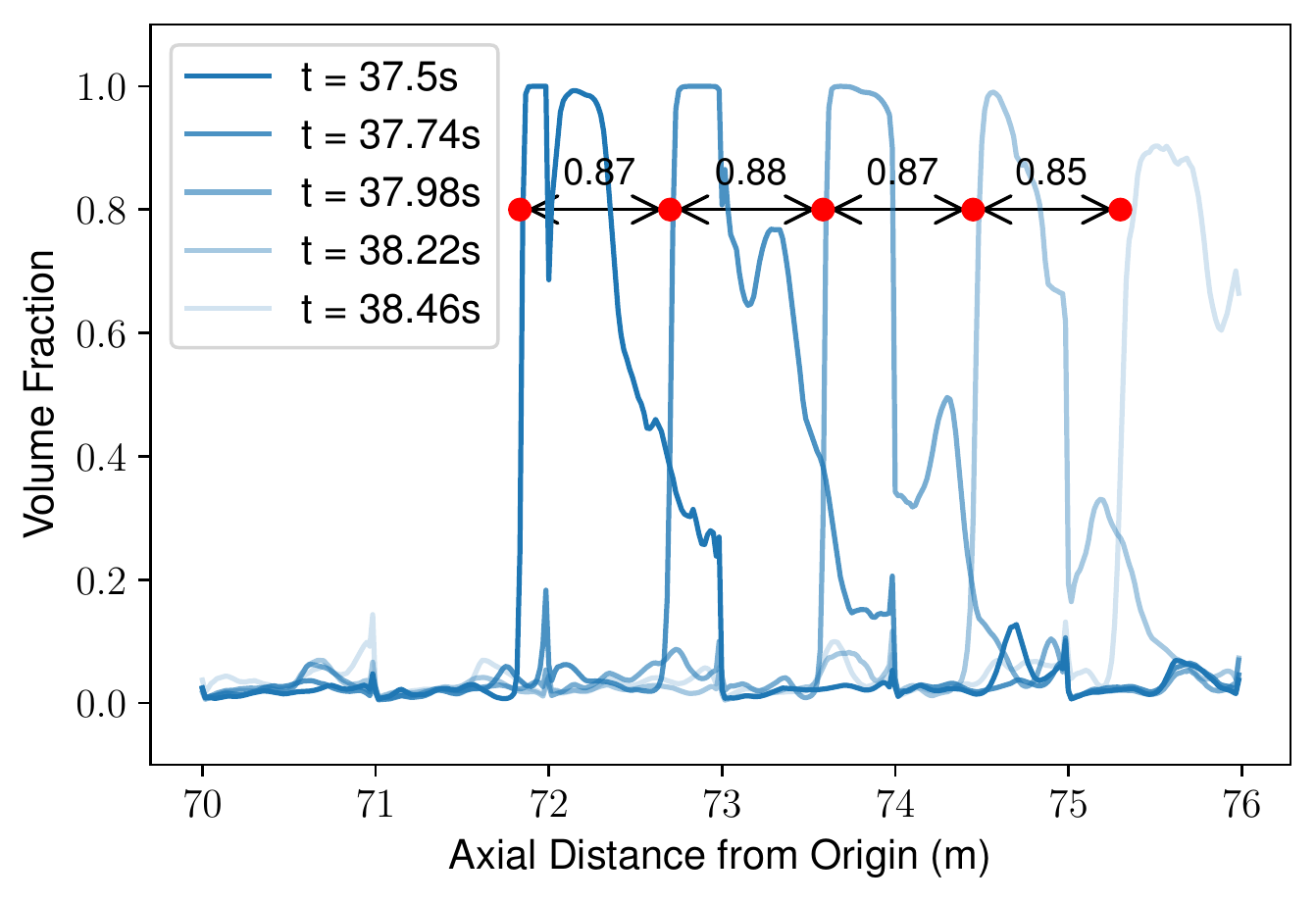}
  \caption{Perturbed instability}
  \label{fig:mass_frac_snapshots_e}
\end{subfigure}
\caption{(a, b) Volume fractions from AI-DDNIROMs with boundary conditions as indicated at a single point in time spanning the domain between 20 to \SI{80}{m} and at a height of \SI{0.0039}{\metre}. (c, d, e) Zoomed in sections from subfigures (a) and (b) and a section of the original simulation, plotted for a few consecutive timesteps.}
\label{fig:mass_frac_snapshots}
\end{figure*}

Figure~\ref{fig:mass_frac_snapshots} follows a pseudo-slug for five time levels as viewed through the volume fraction fields for the original simulation and the reduced-order models. It shows firstly that the instabilities presented in Figures~\ref{fig:mass_frac_snapshots_d} and~\ref{fig:mass_frac_snapshots_e} were similar to an instability that also occurred within the original simulation, presented in Figure~\ref{fig:mass_frac_snapshots_c}.
Furthermore, by observing that the instabilities travelled similar distances between time levels, it can be deduced that they travelled at a similar velocity within the shown timespan as well. While this only presents the dynamics for a single instability at a couple of points in time, the similarity of these situations might reveal that the predictive adversarial model was producing a situation very similar to one it had seen before, which is what this model was hypothesized to do due to its use of the adversarial training strategy.

\begin{figure*}[h!]
\begin{subfigure}{.33\textwidth}
  \centering
  \includegraphics[width=1\linewidth]{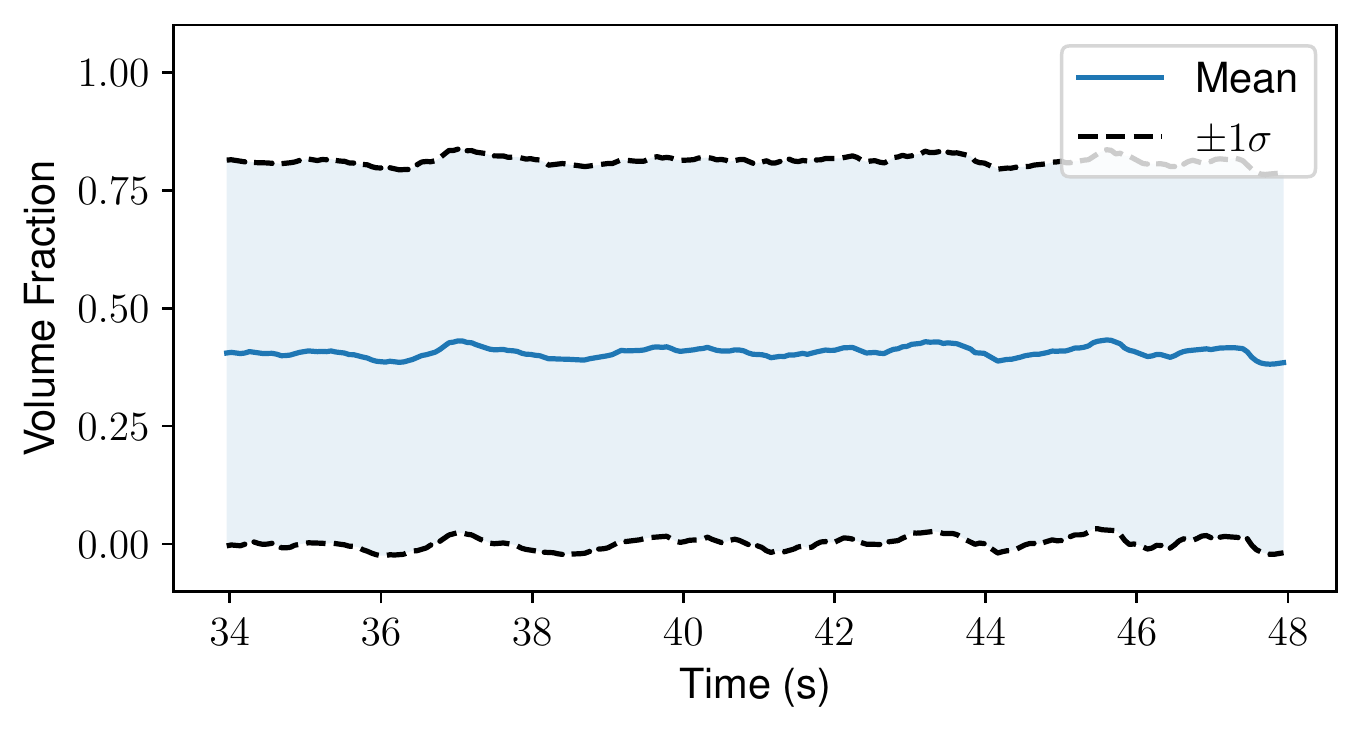}
    \caption{Cycling through slug formation}
  \label{fig:volfrac_avg_pipe_b}
\end{subfigure}
\begin{subfigure}{.33\textwidth}
  \centering
  \includegraphics[width=1\linewidth]{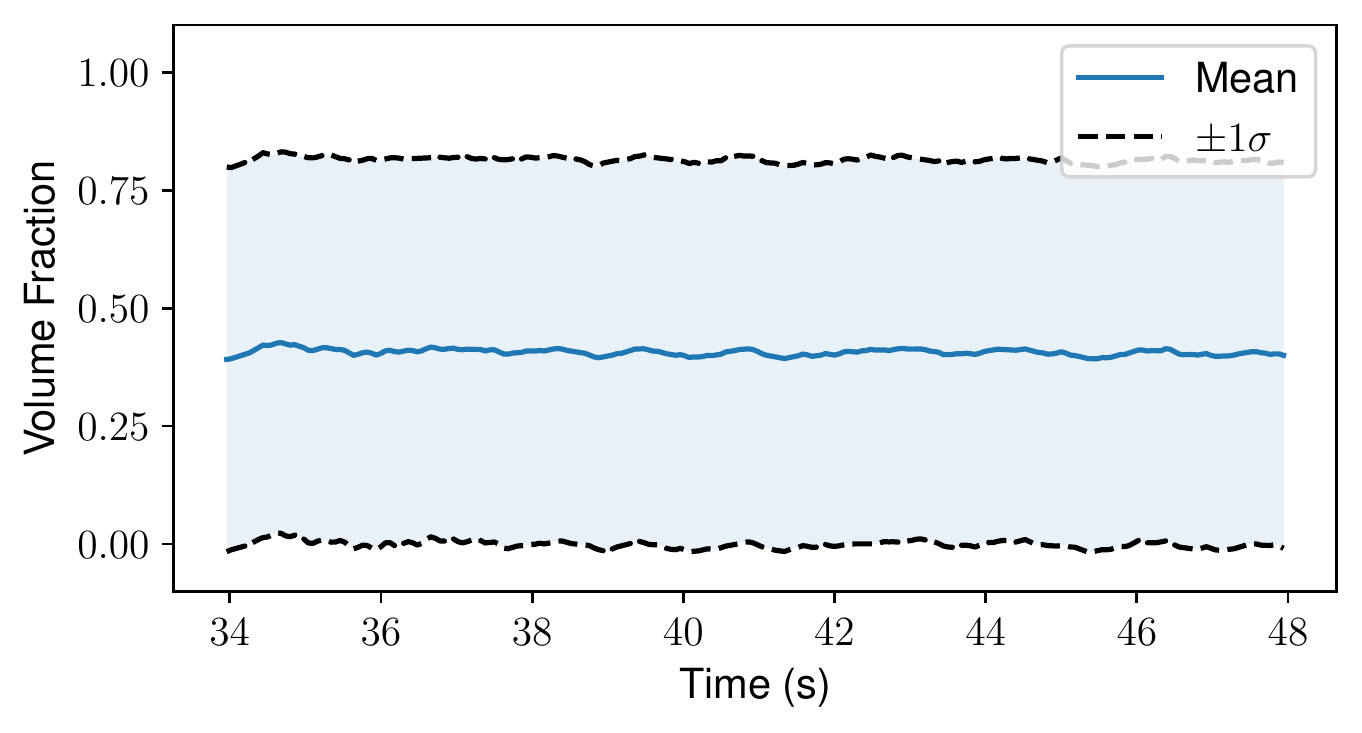}
   \caption{Perturbed instability}
  \label{fig:volfrac_avg_pipe_c}
\end{subfigure}%
\begin{subfigure}{.33\textwidth}
  \centering
  \includegraphics[width=1\linewidth]{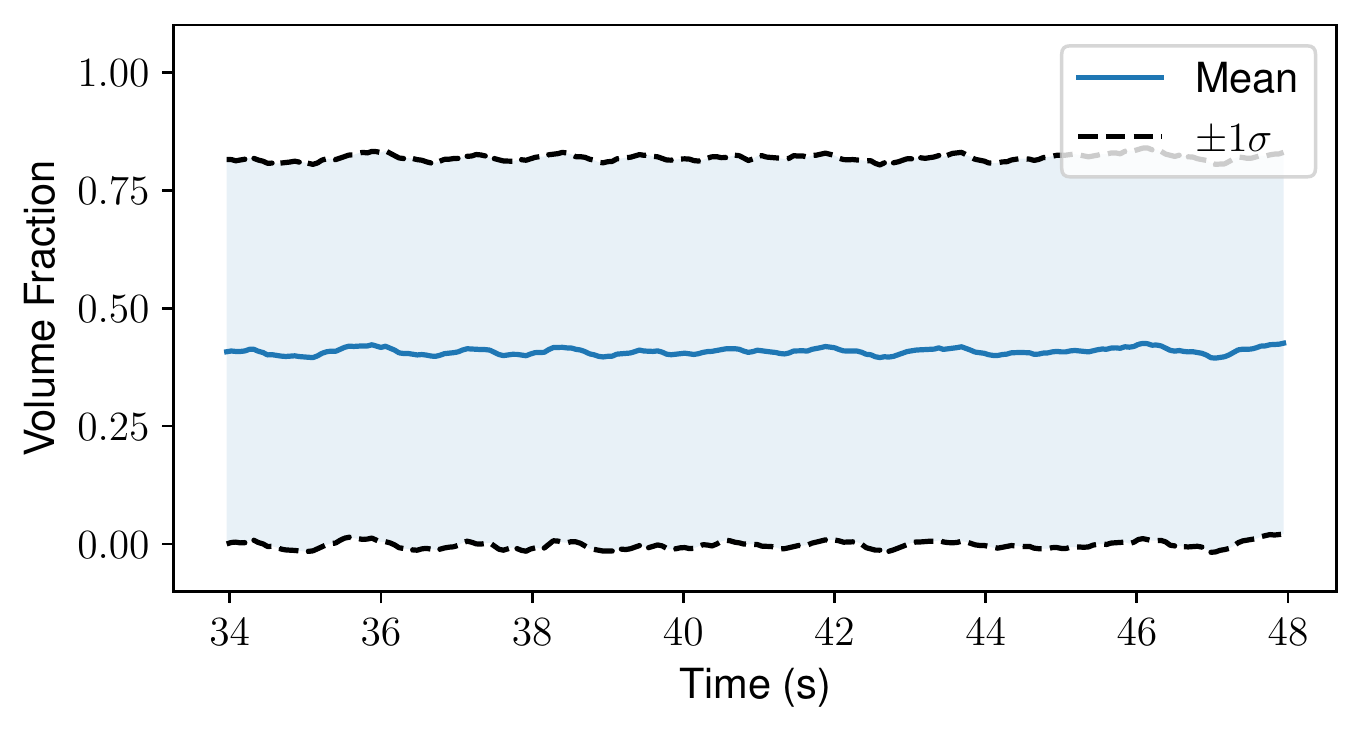}
    \caption{Original boundaries repeated}
  \label{fig:volfrac_avg_pipe_a}
\end{subfigure}
\begin{subfigure}{.33\textwidth}
  \centering
  \includegraphics[width=1\linewidth]{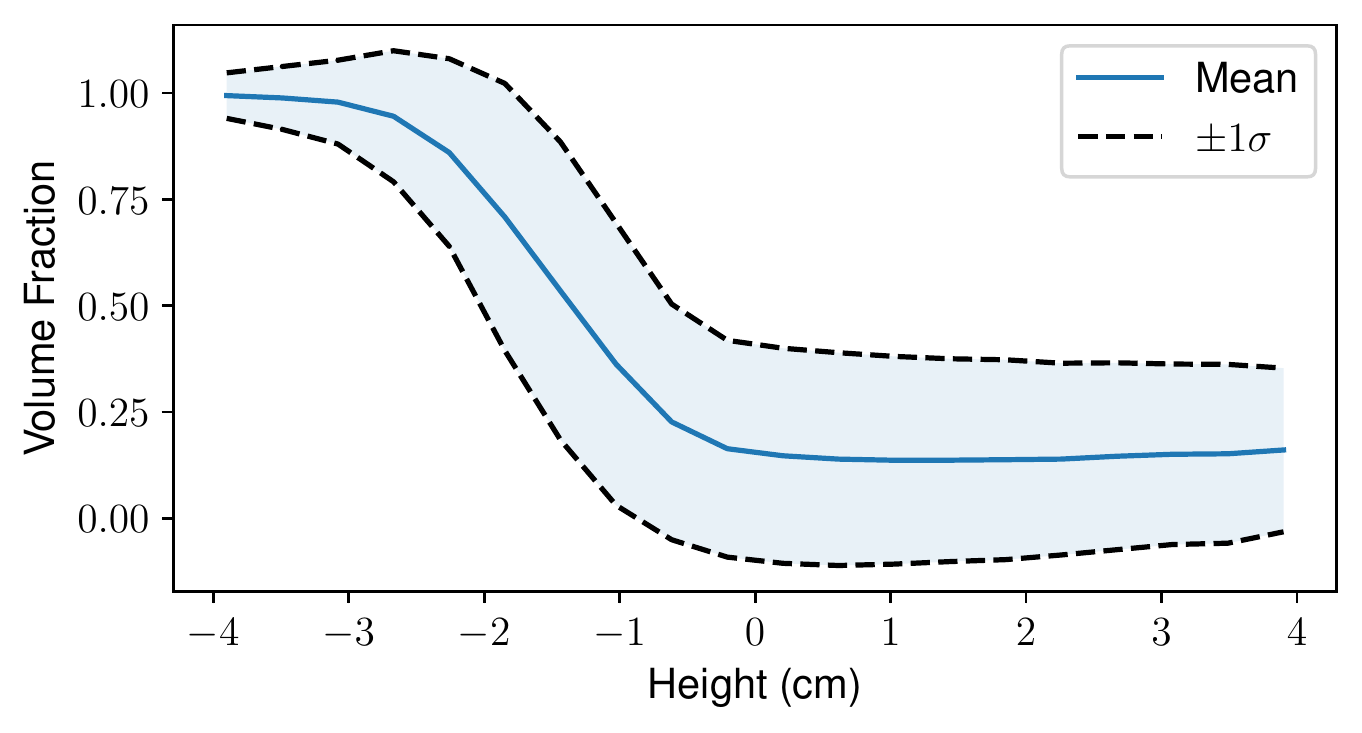}
  \caption{Cycling through slug formation}
  \label{fig:volfrac_avg_pipe_e}
\end{subfigure}%
\begin{subfigure}{.33\textwidth}
  \centering
  \includegraphics[width=1\linewidth]{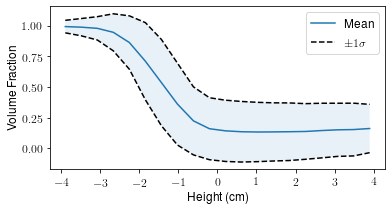}
  \caption{Perturbed instability}
  \label{fig:volfrac_avg_pipe_f}
\end{subfigure}%
\begin{subfigure}{.33\textwidth}
  \centering
  \includegraphics[width=1\linewidth]{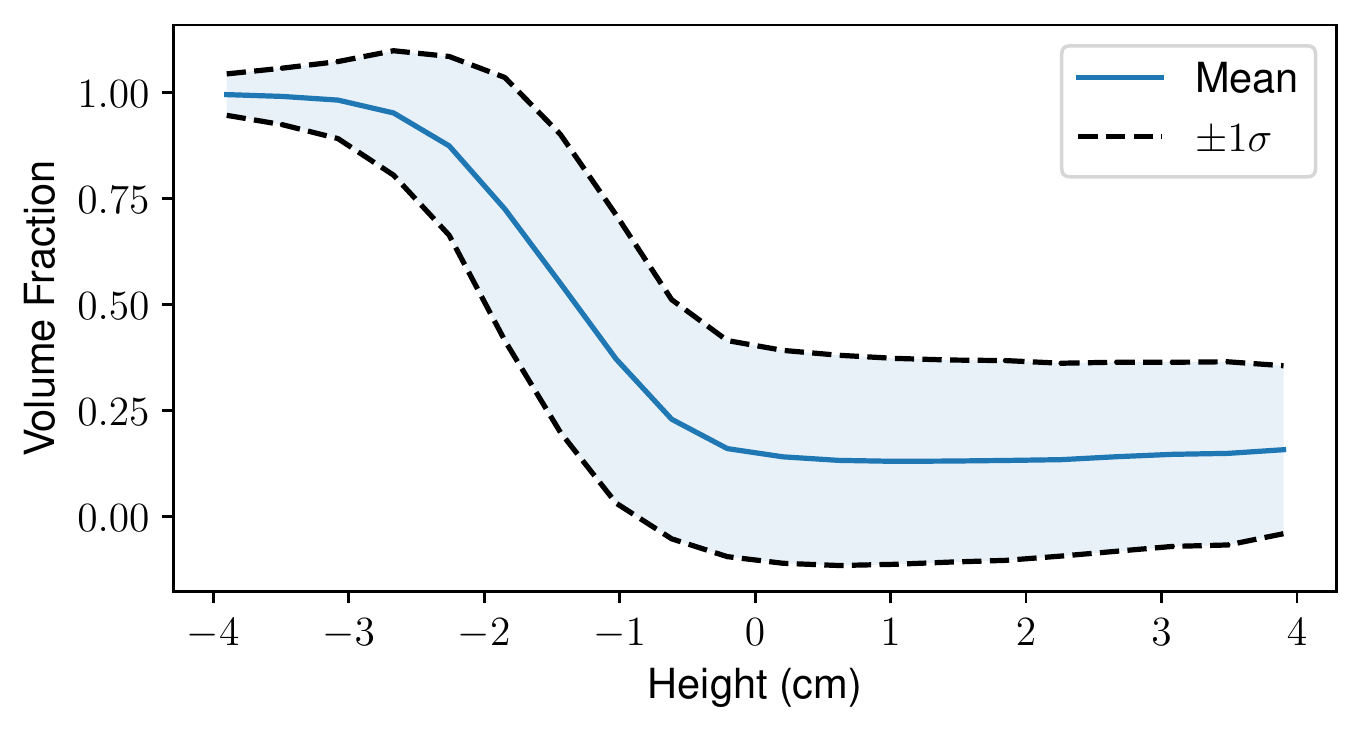}
  \caption{Original boundaries repeated}
  \label{fig:volfrac_avg_pipe_d}
\end{subfigure}
\caption{Mean volume fractions throughout the pipe spanning the domain between \SI{20}{m} and \SI{80}{m} for the different boundary conditions. (a, b, c) Volume fractions averaged over all points in space. (d, e, f) Volume fractions averaged over time, as well as width and length. Note that $\sigma$ refers to the standard deviation.}
\label{fig:volfrac_avg}
\end{figure*}

Figures~\ref{fig:volfrac_avg_pipe_b}, \ref{fig:volfrac_avg_pipe_c} and~\ref{fig:volfrac_avg_pipe_a} show the volume fractions averaged over the full \SI{98}{\metre} domain for a short time period. Note that the start of this time period was chosen so that the influence of the boundaries had already propagated throughout the entire domain. Figures~\ref{fig:volfrac_avg_pipe_e}, \ref{fig:volfrac_avg_pipe_f} and~\ref{fig:volfrac_avg_pipe_d} display the volume fractions averaged over the time period included in the previous three subfigures (\ref{fig:volfrac_avg_pipe_b}-\ref{fig:volfrac_avg_pipe_a}), and the width and length of the domain. These latter three plots thus show how the volume fractions change with respect to height. It is clear from these plots that most of the water collects at the bottom of the pipe. If we assume that the situation in which the original boundaries were repeated in their entirety produced results similar to the original simulation, the fact that the plotted dynamics are similar to the two simulations with artificially generated boundaries suggests strongly that the model was making realistic predictions here for each of the boundaries.

\begin{figure*}[h]
\centering
\begin{subfigure}{.99\textwidth}
  \centering
  \includegraphics[width=1\linewidth]{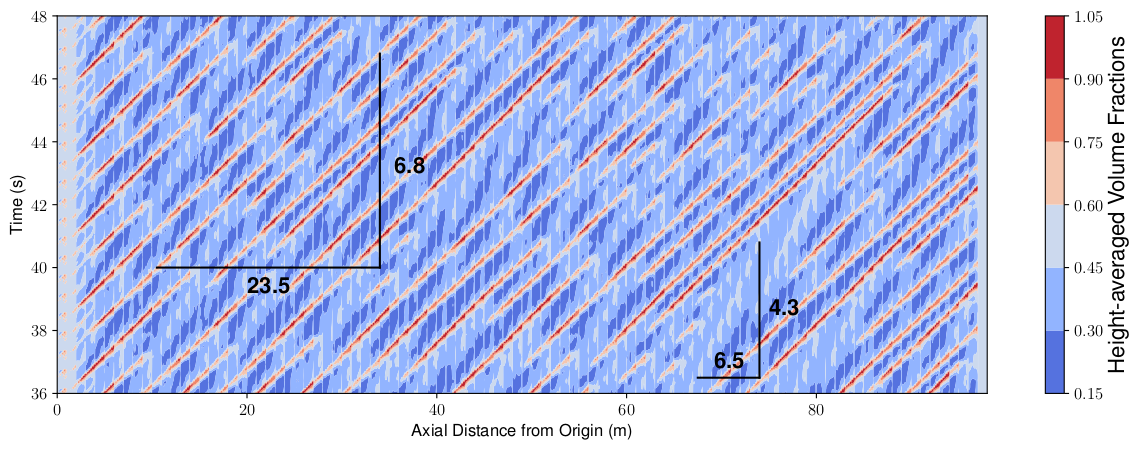}
  \caption{Cycling through slug formation}
  \label{fig:sfig1}
 \end{subfigure}%
 \newline
\begin{subfigure}{.99\textwidth}
  \centering
  \includegraphics[width=1\linewidth]{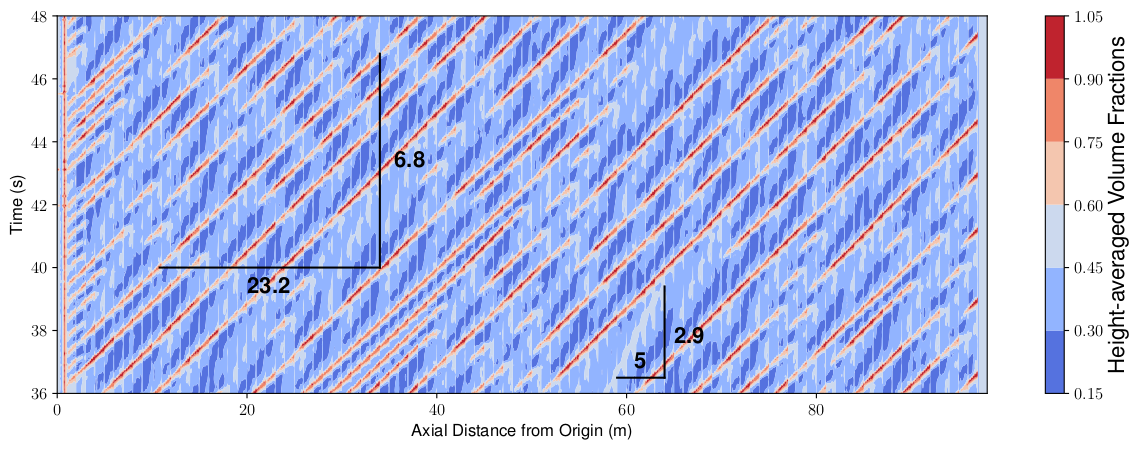}
  \caption{Perturbed instability}
  \label{fig:sfig1}
\end{subfigure}%
 \newline
\begin{subfigure}{.99\textwidth}
  \includegraphics[width=1\linewidth]{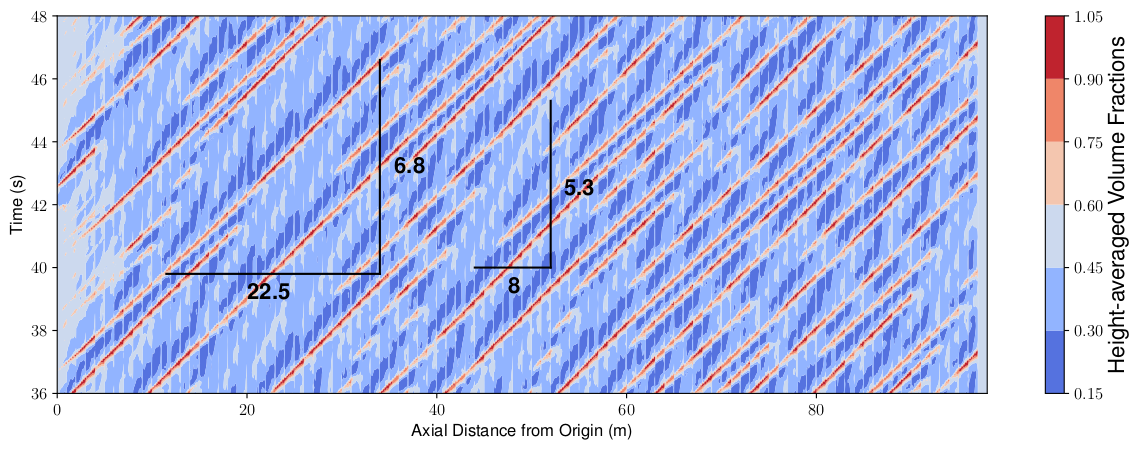}
  \caption{Original boundaries repeated}
  \label{fig:sfig1}
\end{subfigure}
\caption{Volume fractions as a result of the two boundary conditions and original boundaries repeated throughout \SI{12}{\second} in time for the entire \SI{98}{\metre} pipe. Note that the volume fractions were averaged over the entire height of the pipe.}
\label{fig:spacetime_volfrac}
\end{figure*}

Figure \ref{fig:spacetime_volfrac} displays the volume fractions predicted by the AI-DDNIROMs throughout the entire pipe along the centreline ($xz$ plane) for \SI{12}{s} in time and averaged over the height of the pipe. The red bands that stretch diagonally along the plotted domain represent slugs propagating downstream through the domain as time progresses. The slopes of these lines represent the corresponding velocities. Black lines have been drawn on these plots to indicate these liquid slug velocities and also velocities of secondary waves (light blue). The velocity magnitudes are given in Table~\ref{tab:slug_velocities}. From these plots and the table, one can see that both the slug velocities and velocities of the secondary waves produced by different boundary conditions are very similar. Note that the slight variations may have been caused by the interactions of slugs being within each others vicinity. A pattern that is generally observed in each of the three graphs in Figure~\ref{fig:spacetime_volfrac} is that two slugs which are close to one another tend to slowly approach one another. These graphs also clearly display the influence of the boundary at the inlet on the simulation. In fact, the first couple of metres is where the simulations differ the most. However, it seems that after those first couple of metres the simulations all restore to a very similar pattern. 
\begin{table*}[htbp]
\caption{\label{tab:slug_velocities}The slug velocities and secondary wave velocities for the three methods of generating boundary conditions. }
\centering
\begin{tabular}{lp{5mm}cp{5mm}c } 
\toprule
&& slug velocity && secondary wave velocity\\
\toprule
cycling through slug formation&& \SI{3.5}{\metre\per\second} && \SI{1.5}{\metre\per\second}\\
perturbed instability && \SI{3.4}{\metre\per\second} && \SI{1.7}{\metre\per\second}\\
original boundaries repeated && \SI{3.3}{\metre\per\second} && \SI{1.5}{\metre\per\second}\\
\bottomrule
\end{tabular}
\end{table*}


\begin{figure*}[h]
\centering
\begin{subfigure}{.33\textwidth}
  \centering
  \includegraphics[width=1\linewidth]{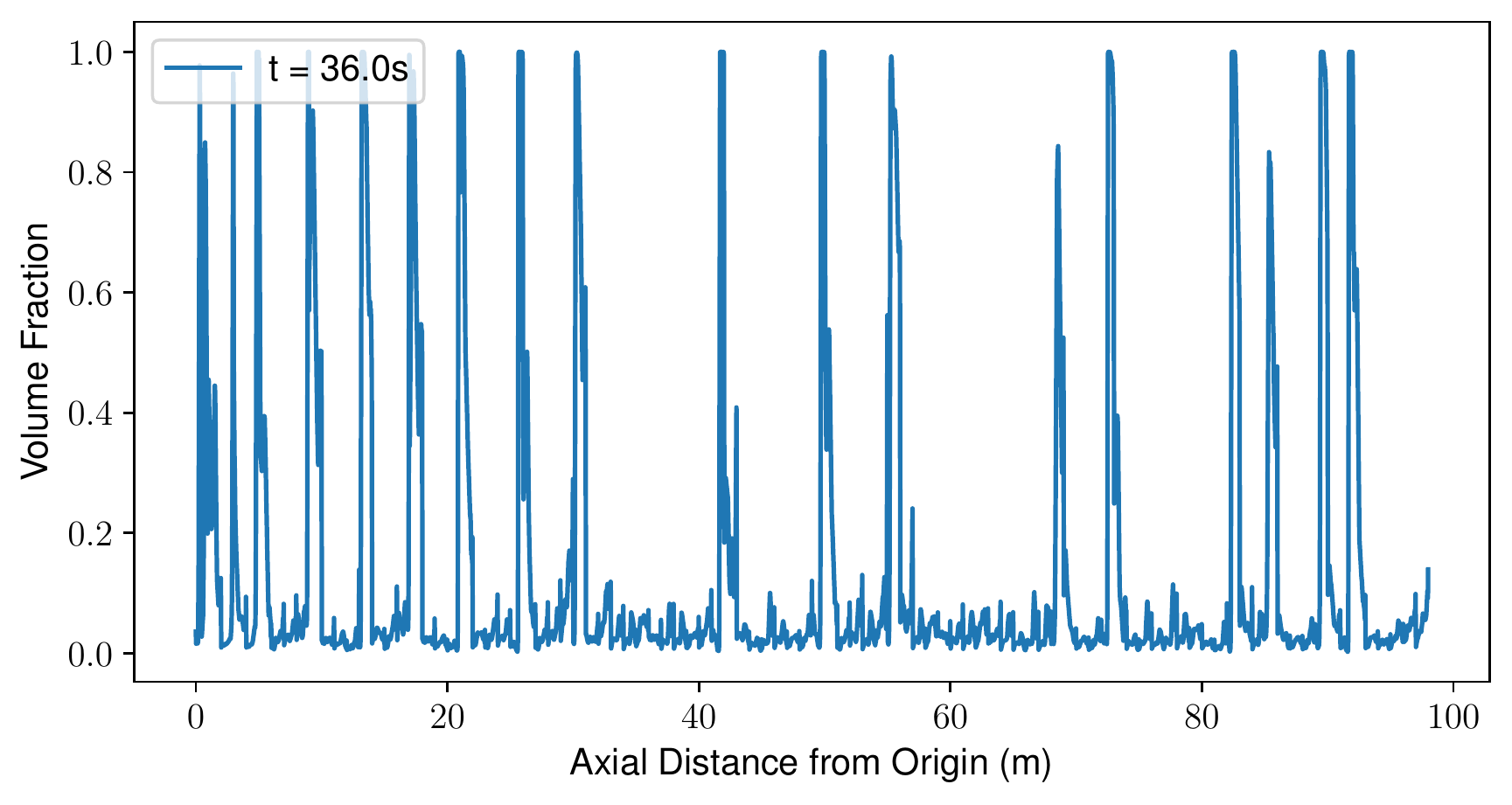}
  \caption{Cycling through slug formation}
  \label{fig:sfig1}
\end{subfigure}%
\begin{subfigure}{.33\textwidth}
  \centering
  \includegraphics[width=1\linewidth]{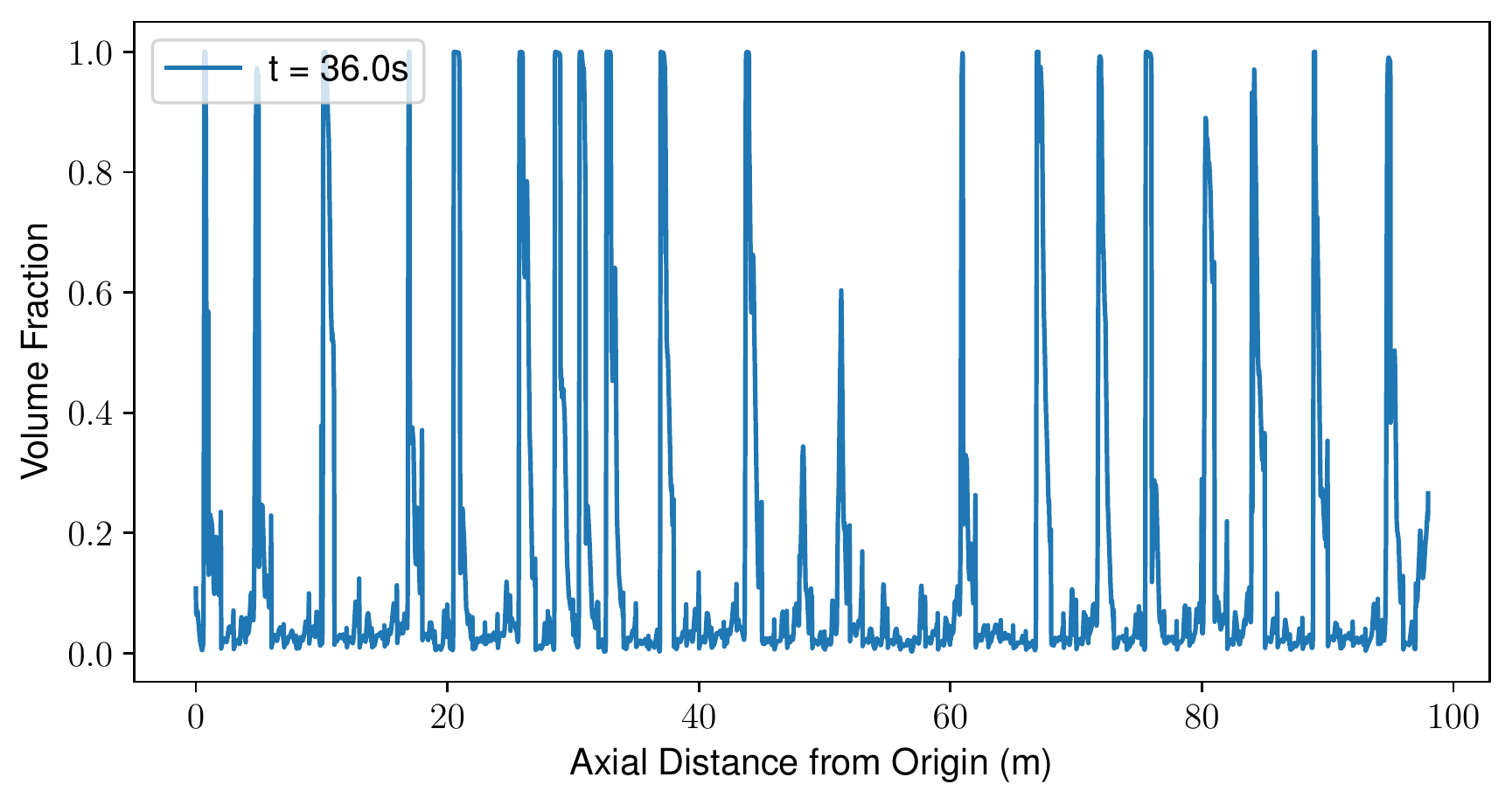}
  \caption{Perturbed instability}
  \label{fig:sfig2}
\end{subfigure}%
\begin{subfigure}{.33\textwidth}
  \centering
  \includegraphics[width=1\linewidth]{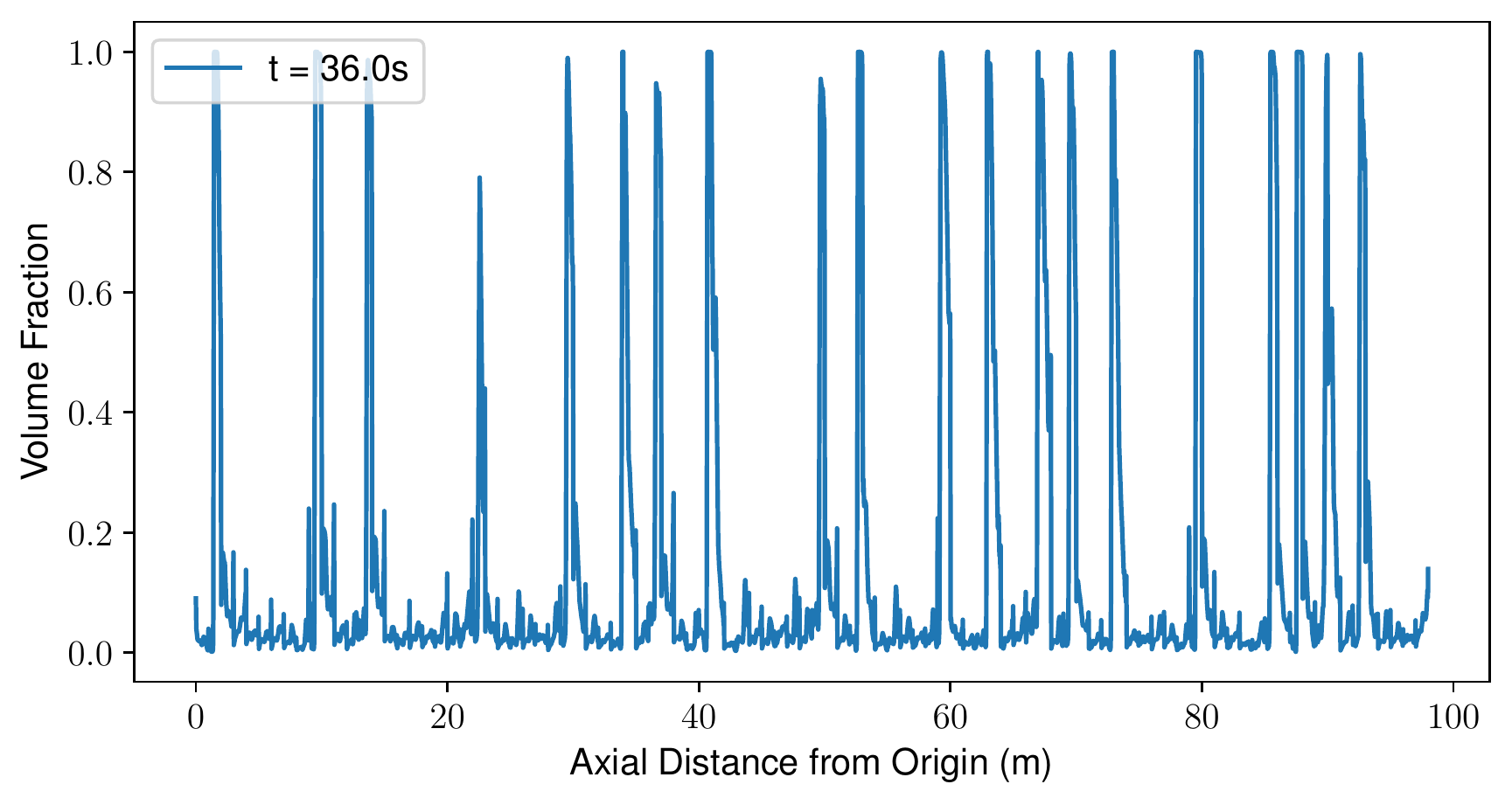}
  \caption{Original boundaries repeated}
  \label{fig:sfig1}
\end{subfigure}
\begin{subfigure}{.33\textwidth}
  \centering
  \includegraphics[width=1\linewidth]{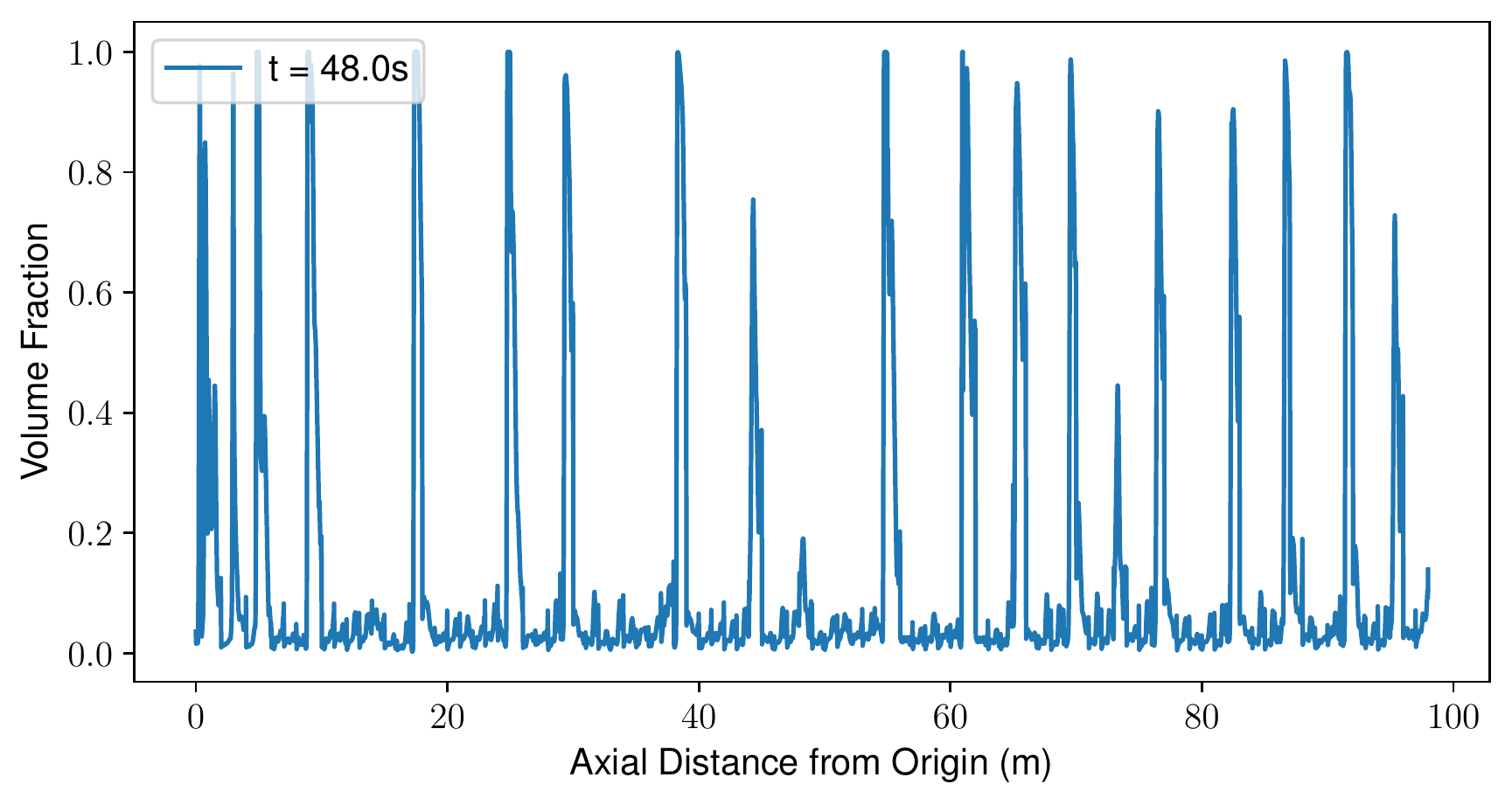}
  \caption{Cycling through slug formation}
  \label{fig:sfig2}
\end{subfigure}%
\begin{subfigure}{.33\textwidth}
  \centering
  \includegraphics[width=1\linewidth]{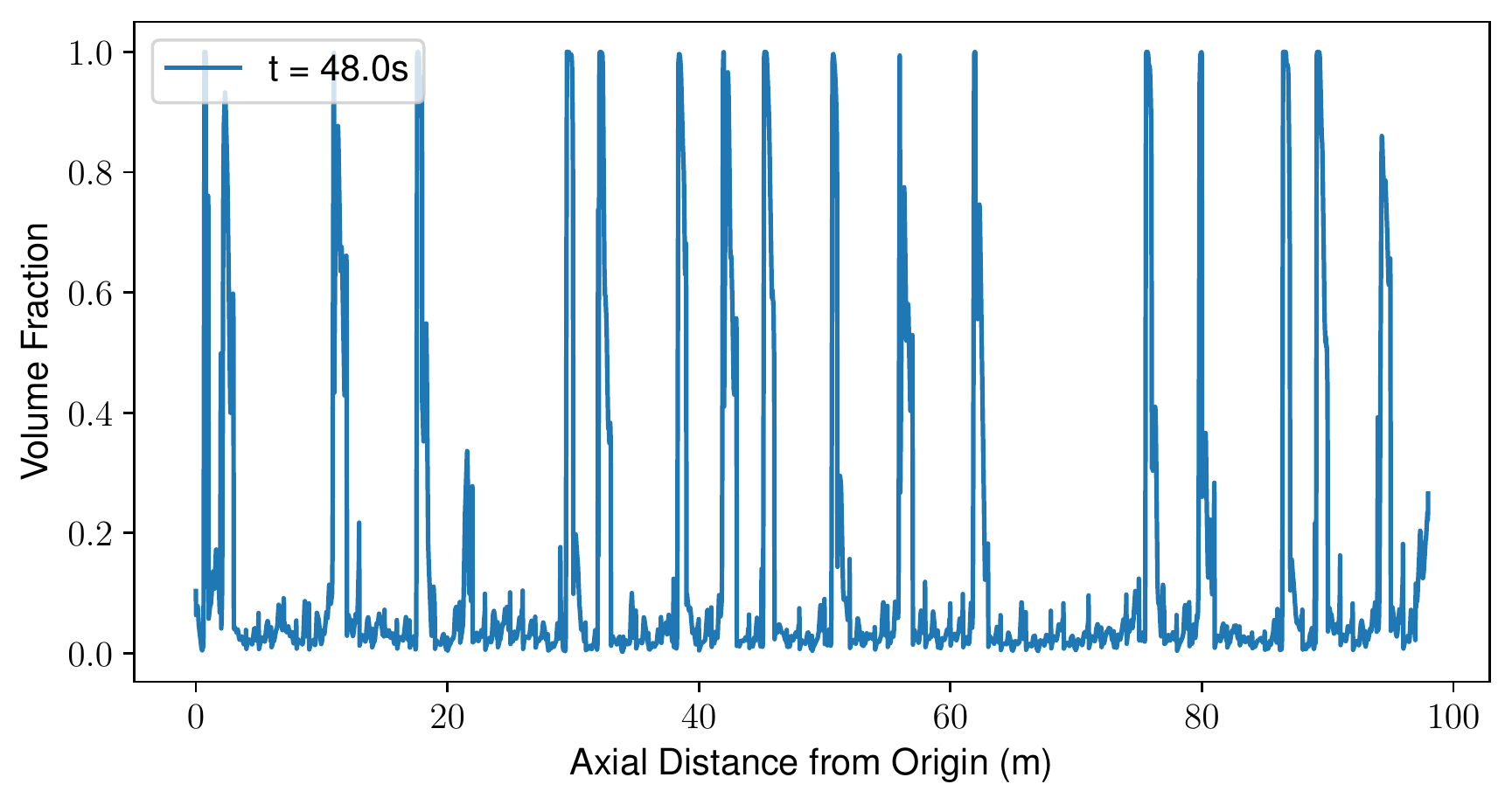}
  \caption{Perturbed instability}
  \label{fig:sfig2}
\end{subfigure}%
\begin{subfigure}{.33\textwidth}
  \centering
  \includegraphics[width=1\linewidth]{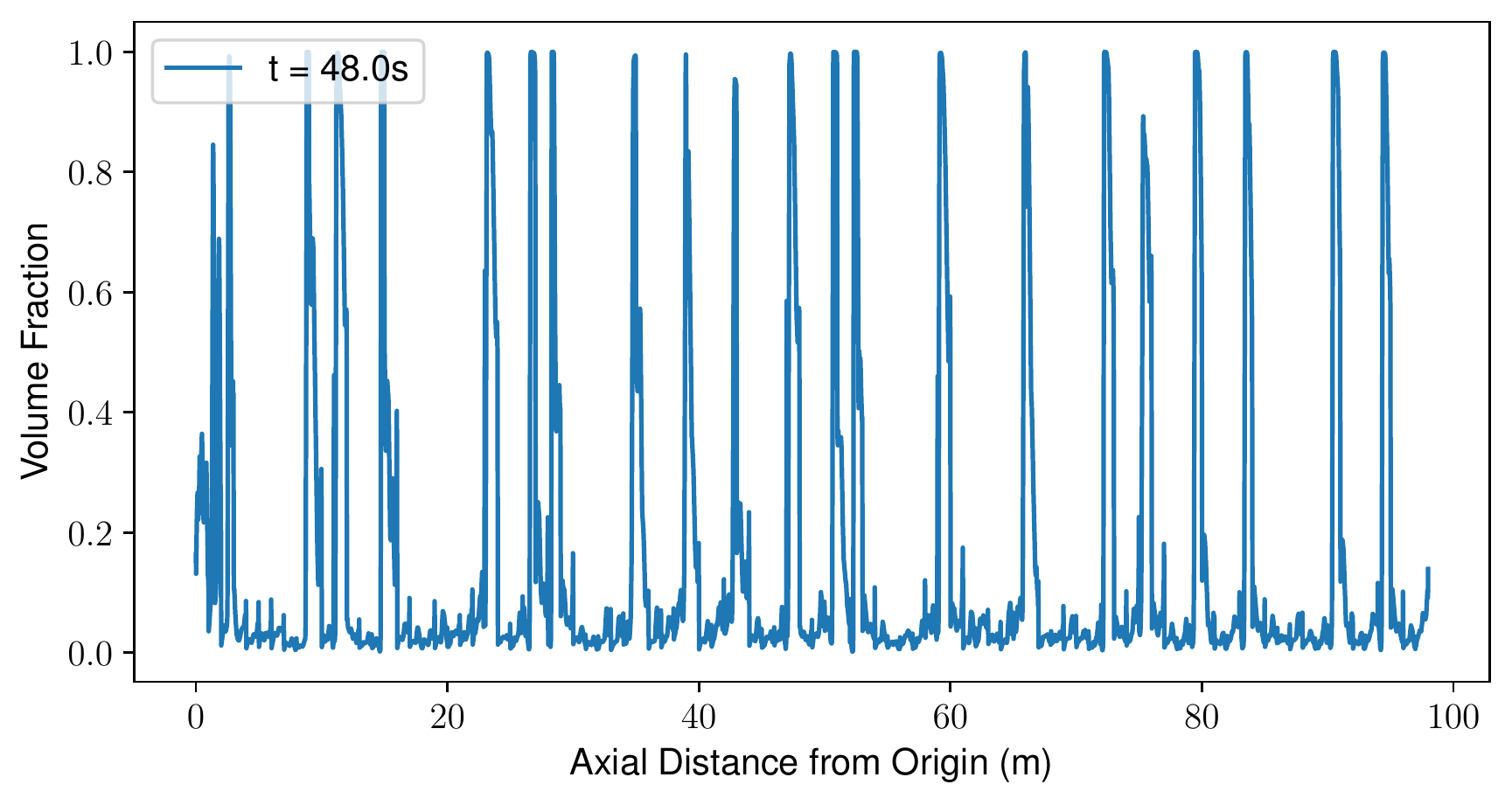}
  \caption{Original boundaries repeated}
  \label{fig:sfig1}
\end{subfigure}
\caption{Volume fractions at \SI{36}{s} (upper) and \SI{48}{s} (lower), at a height of \SI{0.0039}{\metre} above the centreline of the pipe (excluding the first and last subdomain due to the fact that these subdomains showed different values from those throughout the rest of the domain, skewing the vertical axes) for the two predictions with different boundary conditions.}
\label{fig:steps_fulldomain}
\end{figure*}

Figure \ref{fig:steps_fulldomain} shows the volume fractions at a few steps in time to give an impression of these fields throughout the full width of the domain, showing a different perspective of the information seen in Figure~\ref{fig:spacetime_volfrac}.

\subsubsection{Computational times}
The AI-DDNIROM shows a significant computational speed up over the high-fidelity model as expected. The high-fidelity model of the \SI{10}{m} pipe took approximately two weeks to complete (run on processor type Intel\textsuperscript{\scriptsize\textregistered} Xeon\textsuperscript{\scriptsize\textregistered} E5-2640, \SI{2.4}{GHz}), whereas the AA-DDNIROM prediction for the \SI{98}{m} pipe took approximately \SI{20}{minutes} to generate (run on GPUs within Google's Colab platform~\cite{googlecolab,colab2019}).   

\section{Conclusions and Further Work}\label{sec:conclusions}
We present an AI-based non-intrusive reduced-order model combined with a domain decomposition (AI-DDNIROM), which is capable of making predictions for significantly larger domains than the domain used in training the model. The main findings of this study are listed below.
\begin{enumerate}[(i)]
\item For dimensionality reduction, a number of autoencoders are compared with proper orthogonal decomposition, and the convolutional autoencoder is seen to perform the best for both test cases (2D flow past a cylinder and 3D multiphase flow in a pipe). 
\item When training neural networks, it has been observed that computational physics applications typically have access to less training data than image-based applications~\cite{Swischuk2019,Brunton2020} which can lead to poor generalisation. To combat this, for the dimensionality reduction of multiphase flow in a pipe, we use `overlapping' snapshots, that is, in addition to 10 subdomains being equally spaced along the pipe, 10 supplementary subdomains are located at random within the pipe. 
This doubles the amount of training data and results in improved performance.   
\item For prediction we use an predictive adversarial network based on the adversarial autoencoder~\cite{makhzani2015adversarial} but modified to predict in time. This model performs well, gives realistic results, and, unlike feed forward or recurrent networks without such an adversarial layer, does not diverge for the multiphase test case shown here.
\item Finally, we make predictions for a \SI{98}{m} pipe (the `extended pipe') with the AI-DDNIROM that was trained on results from a \SI{10}{m} pipe. Statistics of results from the extended pipe are similar to those of the original pipe, so we conclude that the predictive adversarial network has made realistic predictions for this extended pipe.
\end{enumerate}

A number of improvements could be made to the approach presented here. A physics-informed term could be included in the loss function of either the convolutional autoencoder or the predictive adversarial network. This would ensure that conservation of mass and momentum would be more closely satisfied by the predictions of the neural networks. Secondly, although the initial conditions have little effect on the predictions, the boundary conditions do have a significant effect. Rather than the heuristic approach adopted here, a generative adversarial model (GAN) could be used to predict boundary conditions for the inlet and outlet subdomains. The GAN could be trained to predict the reduced variables at several time levels, then latent variables consistent with all but one of the time levels (the future time level) can be found by an optimisation approach~\cite{Silva2021DA}. From these latent variables, the boundary condition for the future time level can be obtained. %
Finally, a hierarchy of reduced-order models could be used in order to make the approach faster. The lowest-order model could represent the simplest physical features of the flow and the higher-order models could represent more complicated flow features. To decide whether the model being used in a particular subdomain was sufficient, the discriminator of the predictive adversarial network could be used.



\section*{CRediT authorship contribution statement}
\textbf{Claire E. Heaney:} Conceptualization, Methodology, Software, Writing - Original Draft, Writing - Review \& Editing, Supervision; 
\textbf{Zef Wolffs:} Methodology, Software, Writing - Original Draft, Writing - Review \& Editing;
\textbf{J{\'o}n Atli T{\'o}masson:} Software, Writing - Review \& Editing; 
\textbf{Lyes Kahouadji:} Software, Writing - Review \& Editing;
\textbf{Pablo Salinas:} Software, Writing - Review \& Editing;
\textbf{Andr{\'e} Nicolle:} Conceptualization, Writing - Review \& Editing, Supervision; 
\textbf{Ionel M. Navon:} Conceptualization, Writing - Review \& Editing; 
\textbf{Omar K. Matar:} Conceptualization, Writing - Review \& Editing, Supervision, Funding acquisition;
\textbf{Narakorn Srinil:} Conceptualization, Writing - Original Draft, Writing - Review \& Editing, Supervision, Funding acquisition;
\textbf{Christopher C. Pain:} Conceptualization, Methodology, Software, Writing - Original Draft, Writing - Review \& Editing, Supervision, Funding acquisition.

\section*{Declaration of competing interest}
The authors declare that they have no known competing financial interests or personal relationships that could have appeared to influence the work reported in this paper.

\section*{Data availability statement}
The data that support the findings of this study are available from the corresponding author upon reasonable request. Some codes and information about the various neural networks used in this paper can be found in the following Github repository: \url{https://github.com/acse-zrw20/DD-GAN-AE}.

\section*{Acknowledgements}
The authors would like to acknowledge the following EPSRC grants: \href{https://www.muffinsproject.org.uk/}{MUFFINS}, MUltiphase Flow-induced Fluid-flexible structure InteractioN in Subsea applications (EP/P033180/1, EP/P033148/1); RELIANT, Risk EvaLuatIon fAst iNtelligent Tool for COVID19 (EP/V036777/1); the PREMIERE programme grant (EP/T000414/1); MAGIC, Managing Air for Green Inner Cities (EP/N010221/1); and INHALE, Health assessment across biological length scales (EP/T003189/1).

\bibliography{main_elsevier_combined.bib}

\clearpage
\appendix
\section{Hyperparameter optimisation}
Extensive hyperparameter optimsation was carried out for the artificial neural networks used in this investigation. This was done on the \href{https://wandb.ai/site}{Weights \& Biases} platform which allows for efficient searching of high-dimensional parameter space, using methods such as random searches and Bayesian searches. For example, to perform a grid search of the predictive adversarial network for one architecture would involve searching 18 dimensional parameter space, and, with the combinations given in Table~\ref{tab:app_hpo}, would amount to over 2 billion (\num{2e9}) model evaluations (for one architecture). Instead of using a grid search, we perform an initial random search of parameter space, followed by Bayesian optimisation. For the predictive adversarial network this resulted in 1530 model evaluations (for all architectures). The full report for this network is available on \href{https://wandb.ai/zeff020/pred-aae}{Weights \& Biases}.

Table~\ref{tab:app_hpo} shows the range of hyperparameters that were investigated during optimisation for all the networks (the three autoencoder-based networks used for dimensionality reduction for the two test cases and the predictive adversarial network used in multiphase flow in a pipe). These include the exponential decay rate for the first moment estimates ($\beta_1$); the exponential decay rate for the exponentially weighted infinity norm ($\beta_2$); the interval between snapshots (interval) so that an interval of $n$ corresponds to every $n$th snapshot being put in the datasets; the number of discriminator iterations (n discrim); the number of gradient ascent steps (n gradient); 
the standard deviation of the noise that was randomly added to the input of the discriminator within the adversarial autoencoder.
\begin{table*}[htbp]
\caption{\label{tab:app_hpo}Variation and ranges of values studied during the hyperparameter optimisation.} 
\centering
\begin{tabular}{p{0.5cm}lp{12cm}}
\toprule
\multicolumn{3}{l}{\textit{All networks}}\\
&activation functions      & tanh, sigmoid, relu, elu \\
&final activation function & tanh, sigmoid, linear \\
&architecture\footnote{Here a global picture of the architectures is presented, for the source code containing all of the used architectures please see the Github repository: \url{https://github.com/acse-zrw20/DD-GAN-AE/tree/main/ddganAE/architectures}}              & number of layers: 6, $\cdots$, 20 \\
&                           & number of channels: 2, $\cdots$, 128 \\
&                           & dense layer sizes (non-latent): 32, $\cdots$, 2000 \\
&                           & kernel sizes: 3, 5 \\
&                           & layer types: \{1D, 2D, 3D\}-Conv., \{1D, 2D, 3D\}-MaxPool, \{1D, 2D, 3D\}-UpSample, Dense \\
&batch size          &  32, 64, 128 \\
&optimiser           & Adam, Nadam, SGD\\ 
&$\beta_1$  & 0.8, 0.9, 0.98\\
&$\beta_2$           & 0.9, 0.999, 1 \\
&batch normalisation & true, false   \\
&dropout             & 0.3, 0.55, 0.8 \\
&epochs         & 100, 200, 500, 1000, 2000 \\
&interval            & 1, 2, 4, 5, 6, 10\\
&learning rate       & 0.00005, 0.0005, 0.005\\
\multicolumn{3}{l}{\textit{Adversarial networks only}} \\
&discrim architecture$^{\text{a}}$      & number of layers: 3 \\
&                           & dense layer sizes (non-latent): 100, 500, 1000 \\
&                           & layer types: Dense \\
&n discrim           & 1, 2, 5 \\
&n gradient     & 0, 3, 8, 15, 30 $\qquad$\textit{(0 means that no steps of gradient ascent were taken)}\\
&std noise      & 0, 0.00001, 0.001, 0.01, 0.05, 0.1 \\
&regularisation & 0, 0.000001, 0.00001, 0.001\\
\multicolumn{3}{l}{\textit{Predictive Adversarial networks only}} \\
&latent vars         & 30, 50, 100 \\
\bottomrule
\end{tabular}
\end{table*}

Table~\ref{tab:app_optimal_ae} shows the optimal values found in the hyperparameter optimisation for the dimensionality reduction methods based on autoencoders for flow past a cylinder and for multiphase flow in a pipe.

Table~\ref{tab:architecture_DR} gives the optimal architectures found by hyperparameter optimisation the six autoencoder-based networks used in the dimensionality reduction of flow past a cylinder and multiphase flow in a pipe.

Table~\ref{tab:app_optimal_pan} shows the optimal values found in the hyperparameter optimisation for the predictive adversarial network used for the non-intrusive reduced-order model of multiphase flow in a pipe, and Table~\ref{tab:app_optimal_pan_arch} gives the optimal architecture.

\begin{table*}[htbp]
\caption{\label{tab:app_optimal_ae}The optimal values for the hyperparameters of the autoencoders used in the dimensionality reduction stage of flow past a cylinder and  multiphase flow in a pipe.}
\centering
\begin{tabular}{p{2mm}lllllll}
\toprule
& & \multicolumn{3}{c}{\textbf{Flow past a cylinder}} &  \multicolumn{3}{c}{\textbf{Multiphase pipe flow}} \\
& & \textbf{CAE} & \textbf{AAE} & \textbf{SVD-AE} & \textbf{CAE} & \textbf{AAE} & \textbf{SVD-AE} \\
\toprule
\multicolumn{2}{l}{activation functions:}  &&&&&&  \\  
& convolutional layers & elu & elu  & & elu & elu & sigmoid \\
& dense layers        & relu & relu & relu & relu & linear & sigmoid\\
& output layer & & & elu & sigmoid & sigmoid & linear\\
\multicolumn{2}{l}{optimiser:}  &&&&&&  \\  
& method                 & Adam & Nadam & Nadam & Adam & Adam & Nadam\\
& $\beta_1$      & 0.98 & 0.9 & 0.98 & 0.8 & 0.9 & 0.8\\
& $\beta_2$                 & 0.9 & 0.99999 & 0.99999 & 0.9 & 0.9 & 0.99999\\
\multicolumn{2}{l}{batch size} & 128 & 128 & 64 & 64 & 32 & 64\\
\multicolumn{2}{l}{epochs}  & 200 & 200 & 200 & 100 & 1000 & 100 \\
\multicolumn{2}{l}{batch normalisation}       & & & false & & & false\\
\multicolumn{2}{l}{train method}       & & default & & & default &\\
\multicolumn{2}{l}{dropout} & & & 0.55 & & & 0.55\\
\multicolumn{2}{l}{learning rate} & 0.00005 & 0.000005 & 0.0005 & 0.0005 & 0.000005 & 0.00005\\
\multicolumn{2}{l}{regularisation} & 0 & 0 & 0 & 0 & 0 & 0 \\
\bottomrule
\end{tabular}
\end{table*}

\begin{table*}[htbp]
\renewcommand{\arraystretch}{0.7}
\caption{\label{tab:architecture_DR}\linespread{1.1}\selectfont{}The optimal architectures of the autoencoder-based networks used for dimensionality reduction. The figures in the table are the dimensions of the outputs of each layer. For tuples, the final value represents the number of channels or feature maps. The layer type can be convolutional (Conv), maxpooling (MaxPool) or upsampling (UpSample). Flatten layers take an $n$-dimensional array as an input and return a 1D array as an output. A reshape layer converts a 1D input to have the indicated output dimensions. ${}^\dagger$ denotes a layer which is followed by a dropout layer during training. ${}^\ddagger$ denotes convolutional layers which have no padding. In all other cases padding is set so that the output has the same dimensions as the input array, although the number of channels may vary. }
\centering
\begin{tabular}{lcccccc}
\toprule
       & \multicolumn{3}{c}{\textbf{Flow past a cylinder}} & \multicolumn{3}{c}{\textbf{Multiphase flow in a pipe}}\\
layers & \multicolumn{1}{c}{\textbf{CAE}} & \multicolumn{1}{c}{\textbf{AAE}} & \multicolumn{1}{c}{\textbf{SVD-AE}} &  \multicolumn{1}{c}{\textbf{CAE}} & \multicolumn{1}{c}{\textbf{AAE}} & \multicolumn{1}{c}{\textbf{SVD-AE}}\\
\toprule
input & (55,42,2)  & (55,42,2)   & 100 & (60,20,20,4) & (60,20,20,4) & 100\\
Conv & (55,42,32)      & (55,42,32)  &   &  (60,20,20,32) & (60,20,20,32) \\
MaxPool & (28,21,32)     & (28,21,32)  &   & (30,10,10,32) & (30,10,10,32) & \\
Conv & (28,21,64)      & (28,21,64)  &   & (30,10,10,64) & (30,10,10,64) & \\
MaxPool & (14,11,64)     & (14,11,64)  &   & (15,5,5,64) & (15,5,5,64) & \\
Conv & (14,11,128)     &   ---      &    & (15,5,5,128) & (15,5,5,128) & \\
MaxPool & (7,6,128)      &  ---       &    & (8,3,3,128) &  (8,3,3,128)   & \\
\midrule
flatten & 5376           &  9856       &    ---             & 9216  & 9216  & ---\\
Dense 1 &  2688          &  9856       &  500$^{\,\dagger}$ & 10    &  4608 & 1500\\
Dense 2 &    10          &  4926       &  500$^{\,\dagger}$ & 9216  &  10 & 2000\\
Dense 3 &  2688          &    10       &  10               &  ---   & 4608 & 10\\
Dense 4 &  5376     &  4926       &  500$^{\,\dagger}$ & ---   &  9218 & 1500\\
Dense 5 &  ---            &  9856       &  500$^{\,\dagger}$ & --- & --- & 2000\\
Dense 6 &  ---            &  9856       &   ---              & --- & --- & ---\\
\midrule
reshape & (7,6,128)      & (14,11,64)  &  & (8,3,3,128) & (8,3,3,128)  &   \\
Conv & (7,6,128)       & (14,11,64)  &  & (8,3,3,128) &   (8,3,3,128)  &   \\
UpSample & (14,12,128)   & (28,22,64)  &  & (16,6,6,128) & (16,6,6,128)  &   \\
Conv & (14,12,64)      & (28,22,32)  &  & (16,6,6,64) &  (16,6,6,64)  &   \\
Upsample & (28,24,64)    & (56,44,32)  &  & (32,12,12,64) & (32,12,12,64)  &  \\
Conv & (28,24,32)      & (56,44,2)   &  & (30,10,10,32)$^\ddagger$ & (30,10,10,32)$^\ddagger$  & \\
UpSample & (56,48,32)    & ---         &  & (60,20,20,32) & (60,20,20,32) &\\
Conv & (56,48,2)       & ---         &  & (60,20,20,4) & (60,20,20,4)&\\
crop & (55,42,2)         & ---   &  & --- & ---&\\
output & ---        & ---          & 100 &  --- & --- & 100\\
\midrule
trainable & \num{29300300} & \num{291587010} & \num{612110} & \num{1196238} & \num{86001860} & \num{6392110} \\
parameters & & & & & & \\
\bottomrule
\end{tabular}
\end{table*}

\begin{table}[htbp]
\caption{\label{tab:app_optimal_pan}The optimal values of the hyperparameters for the predictive adversarial network found by optimisation for multiphase flow in a pipe.}
\centering
\begin{tabular}{p{4mm}llp{2.5cm}ll}
\toprule
\multicolumn{2}{l}{activation functions:}      &       & & dropout  & 0.3\\
&convolutional layers$\quad$  & relu                   &    & interval &6\\
&dense layers      & relu                             &  & learning rate &0.00005\\
&final layer       & tanh                              & & latent vars &100\\
\multicolumn{2}{l}{optimiser:} &    &                    & n discrim &1\\
&method                   & Nadam                    &   & n gradient &15\\
&$\beta_1$      & 0.98                                &  & std noise&0.01\\
&$\beta_2$                 & 0.9                       & & regularisation&0.001\\
\multicolumn{2}{l}{batch size}                & 32    &  & batch normalisation$\quad$& true\\
\multicolumn{2}{l}{epochs}                    & 2000    &&  training method&weighted loss\\
\bottomrule
\end{tabular}
\end{table}

\begin{table}[htbp]
\caption{\label{tab:app_optimal_pan_arch}The optimal architecture of the predictive adversarial network found by hyperparameter optimisation for multiphase flow in a pipe.}
\centering
\begin{tabular}{lcc}
\toprule
    layers & Predictive Adversarial Network & Discriminator \\
\toprule
input & 30 & 100  \\
Dense 1 & 500 & 100 \\ 
Dense 2 & 500 & 500\\
Dense 3 & 100 &\\
Dense 4 & 500 &\\
Dense 5 & 500 & \\
output  & 10  & 1 \\
\midrule
trainable & \num{622110} & \num{61101} \\
parameters & & \\
\bottomrule
\end{tabular}
\end{table}


\end{document}